\documentclass[preprint,10pt]{elsarticle}
\usepackage{xargs,enumitem}
\usepackage{graphicx, epsfig, caption}
\usepackage{amsfonts, amsmath, amssymb, amsthm}
\usepackage{makecell}
\usepackage{algorithm, algcompatible, verbatim}
\usepackage{lineno}
\usepackage[usenames, dvipsnames]{color}
\usepackage[pdftex,dvipsnames]{xcolor}  
\usepackage[colorinlistoftodos,prependcaption,textsize=tiny]{todonotes}
\usepackage{physics}
\usepackage{subcaption}

\usepackage{algorithm}
\usepackage{algpseudocode}
\algnewcommand\INPUT{\item[\textbf{Input:}]}
\algnewcommand\OUTPUT{\item[\textbf{Output:}]}

\graphicspath{./}

\usepackage[normalem]{ulem}

\begin{document}

\begin{frontmatter}

\title{Quadratic Approximation Manifold for Mitigating the Kolmogorov Barrier in Nonlinear Projection-Based Model Order Reduction}

\author[label1]{Joshua Barnett}
\author[label1,label2,label3]{Charbel Farhat}
  
\address[label1]{Department of Mechanical Engineering, Stanford University, Stanford, CA 94305}
\address[label2]{Department of Aeronautics and Astronautics, Stanford University, Stanford, CA 94305}
\address[label3]{Institute for Computational and Mathematical Engineering, Stanford University, Stanford, CA 94305}

\begin{abstract}
A quadratic approximation manifold is presented for performing nonlinear, projection-based, model order reduction (PMOR). It constitutes a departure from the traditional
affine subspace approximation that is aimed at mitigating the Kolmogorov barrier for nonlinear PMOR, particularly for convection-dominated transport problems. It builds on the data-driven 
approach underlying the traditional construction of projection-based reduced-order models (PROMs); is application-independent; is linearization-free; and therefore is robust for highly nonlinear problems.
Most importantly, this approximation leads to quadratic PROMs that deliver the same accuracy as their traditional counterparts using however a much smaller dimension -- typically, 
$n_2 \sim \sqrt n_1$, where $n_2$ and $n_1$ denote the dimensions of the quadratic and traditional PROMs, respectively. The computational advantages of the proposed high-order approach to nonlinear
PMOR over the traditional approach are highlighted for the detached-eddy simulation-based prediction of the Ahmed body turbulent wake flow, which is a popular CFD benchmark problem in the automotive industry. For a fixed
accuracy level, these advantages include: a reduction of the total offline computational cost by a factor greater than five; a reduction of its online wall clock time 
by a factor greater than 32; and a reduction of the wall clock time of the underlying high-dimensional model by a factor greater than two orders of magnitude.

\end{abstract}


\end{frontmatter}


\section{Introduction}
\label{sec:INTRO}

Projection-based model order reduction (PMOR) is a mathematical technique for reducing the dimensionality $N$ of a high-dimensional model (HDM) of interest through a subspace approximation
$\tilde {\mathbf{u}}$ of the solution $\mathbf{u} \in \mathbb{R}^N$ associated with the HDM. Typically, $\tilde{\mathbf{u}} \in \mathbb{R}^N$ is an affine approximation of the form 
$\tilde {\mathbf{u}} = \mathbf{V} \mathbf{q} + \mathbf{u}_{\text{ref}}$, where $\mathbf{V}\in \mathbb{R}^{N \times n}$ is referred to in general as a {\it right} reduced-order basis (ROB),  $n \ll N$,
$\mathbf{q} \in \mathbb{R}^n$ is known as the vector of generalized coordinates,
and $\mathbf{u}_{\text{ref}} \in \mathbb{R}^N$ is a reference solution.  Using this approximation and a projection that may or may not be orthogonal, PMOR transforms the HDM of interest into a lower dimensional
computational model of dimension $n \ll N$ known as a projection-based reduced-order model (PROM). The technique is becoming increasingly invaluable for many different forms of {\it parametric} 
applications arising in computational structural dynamics (CSD)~\cite{besselink2013comparison, farhat2014dimensional}, 
multiscale modeling~\cite{hernandez2014high, ghasemi2015fast, ohlberger2016model, zahr2017multilevel}, computational fluid dynamics 
(CFD)~\cite{washabaugh2012nonlinear, crowell2012model, carlberg2013gnat}, uncertainty quantification (UQ)~\cite{soize2017nonparametric}, model predictive control 
(MPC)~\cite{ou2009model}, optimization~\cite{yoon2010structural}, and multidisciplinary design analysis and optimization 
(MDAO)~\cite{boncoraglio2021model}. It allows for the parsimonious representation of a large-scale, dynamically complex, and computationally intensive HDM. 
Hence, it is crucial for computations where real-time results are desired.

In the parlance of finite element (FE) modeling, PMOR is essentially a semi-discretization method with {\it global} shape and test functions. Given a parameterization of the HDM in a parameter
space $\mathcal D$, the shape functions are constructed \textit{a posteriori} -- that is, after some knowledge of the application of interest is discovered. The discovery is achieved by sampling 
$\mathcal D$ at some carefully selected parameter points and exercising the HDM at each sampled parameter point. Specifically, one or several solution snapshots are computed at each 
sampled parameter point and collected in a matrix $\mathbf{S} \in \mathbb{R}^{N \times N_s}$, where $N_s$ denotes the total number of solution snapshots. Then, the matrix $\mathbf{S}$ is
{\it compressed} to construct the ROB $\mathbf{V}\in \mathbb{R}^{N \times n}$, where $n \le N_s$, and often $n \ll N_s$. For example, the compression can be obtained by performing the {\it thin} 
singular value decomposition (SVD) of $\mathbf{S}$ and truncating the orthogonal matrix (with respect to some metric) $\mathbf{U} \in \mathbb{R}^{N \times N}$ associated with the nonzero left singular 
values of $\mathbf{S}$ -- which spans the range of $\mathbf{S}$ -- to the low dimension $n \ll N$: this sample compression approach leads to a PMOR method that is equivalent to the method of snapshots 
for proper orthogonal decomposition (POD)~\cite{sirovich1987turbulence}. As for the global test functions, they can be chosen to be identical to the global shape functions, which leads to the
construction of a {\it left} or {\it test} ROB $\mathbf{W} = \mathbf{V}$ and that of a Galerkin PMOR method. Alternatively, the test functions can be constructed such that 
$\mathbf{W} \in \mathbb{R}^{N\times n}$ but $\mathbf{W} \ne \mathbf{V}$, in which case the resulting PMOR method is known as a Petrov-Galerkin PMOR method. For all these reasons, PMOR methods
can be considered to be {\it simulation-driven}, or {\it data-driven} computational methods.

It follows that as discussed in~\cite{grimberg2020stability}, PMOR should benefit from the rich history of development of semi-discretization methods. Hence, for elliptic (e.g. structural mechanics), 
parabolic (e.g. heat conduction), and second-order hyperbolic (e.g. second-order dynamical systems such as those arising in wave propagation and structural dynamics) partial differential equations
(PDE)s, Galerkin PMOR is appropriate. For first-order hyperoblic (e.g. convection-dominated turbulent flow) PDEs however, Petrov-Galerkin PMOR is more appropriate if not 
essential~\cite{grimberg2020stability}.  

Galerkin PMOR is matured for both linear and nonlinear problems, from all theoretical, algorithmic, and application viewpoints~\cite{antoulas2000survey}. In comparison, Petrov-Galerkin PMOR is less 
advanced, although the benefits of Petrov-Galerkin projection to numerical stability have been well established theoretically for linearized first-order hyperoblic 
problems~\cite{amsallem2012stabilization}; and numerically for nonlinear, convection-dominated, turbulent flow problems~\cite{grimberg2020stability}. For the latter problems, the state of the art is 
currently represented by the least squares Petrov-Galerkin (LSPG) method first developed in~\cite{carlberg2011efficient} under the abbreviation GNAT (Gauss-Newton method with Approximated Tensors).
The performance of LSPG in terms of robustness, accuracy, dimensionality reduction, and most importantly, wall clock time reduction, has been most recently demonstrated in~\cite{grimberg2021mesh}, for a 
very large-scale convection-dominated Reynolds-Averaged Navier-Stokes (RANS) application featuring a complex geometry; and in~\cite{grimberg2020stability}, for the Large Eddy Simulation (LES) of a 
compressible turbulent flow.

During the last decade, nonlinear PMOR in general has benefited from parallel advancements in hyperreduction methods -- that is, numerical approaches that approximate the computation of projected 
quantities to ensure a computational complexity that does not scale with the large dimension $N$ of the underlying HDM. Specifically, the state of the art of the empirical interpolation
method (EIM)~\cite{barrault2004empirical} and its discrete counterpart (DEIM)~\cite{chaturantabut2010nonlinear} has been advanced (for example, see ~\cite{antil2014application, hesthaven2016empirical}); 
and the energy-conserving sampling and weighting (ECSW) hyperreduction method originally developed in~\cite{farhat2014dimensional} for the Galerkin PMOR of second-order dynamical systems has been 
extended in~\cite{farhat2019feasible} to linear stochastic PROMs, and in~\cite{grimberg2021mesh}, to nonlinear Petrov-Galerkin PROMs of first-order hyperbolic problems such as convection-dominated 
turbulent flow problems.

Despite the aforementioned advancements, the Kolmogorov barrier remains an outstanding issue for nonlinear PMOR, due to the common reliance of traditional PMOR on an affine subspace approximation.
The barrier is suggested by the slow decay of the Kolmogorov $n$-width~\cite{pinkus1985} $d_n\left(\mathcal{M}\right)$ associated with a subset $\mathcal M$ of a normed space. Specifically, 
$d_n\left(\mathcal{M}\right)$ defines the worst-case error arising from the projection of points in $\mathcal M$ onto the best-possible {\it linear} space of small dimension $n \ll N$, represented
in the case of PMOR by the right ROB $\mathbf{V} \in \mathbb{R}^{N\times n}$. For linear PDEs, $d_n\left(\mathcal{M}\right)$ decays as an exponential function of $n$, which allows achieving a 
minuscule projection error for even a modest size of $n$. For highly nonlinear PDEs however, the decay of $d_n(\mathcal{M})$ with $n$ is significantly slower, which is why the $n$-width is 
often referred to in the literature as a barrier to reducibility. For example, for convection-dominated PDEs, $d_n(\mathcal M) = \mathcal{O}(n^{-1/2})$~\cite{greif2019decay}.

In general, the number of points to sample in $\mathcal D$ in order to construct the right ROB $V \in \mathbb{R}^{N\times n}$ -- and therefore, the number of solution snapshots $N_s$ -- grows 
exponentially with the dimension $N_{\mathcal D}$ of the parameter space $\mathcal D$. Hence, while the $n$-width issue outlined above is independent of any parametric setting, it can be significantly 
exacerbated in the presence of a high-dimensional parametric setting, which in the case of nonlinear, convection-dominated transport problems may lead to an unaffordable ROB $\mathbf{V}$ and associated 
PROM.  Consequently, many attempts to overcome, or at least mitigate, the Kolmogorov barrier have been made in the literature. Most of them amount to abandoning the affine (or linear) subspace 
approximation in favor of some form of nonlinear approximation. In the context of nonlinear, convection-dominated transport problems, these attempts include: the PMOR method based on most-appropriate 
local right ROBs~\cite{amsallem2012nonlinear}, which essentially constructs a piece-wise affine approximation $\tilde{\mathbf{u}}$ of the solution; PMOR approaches based on autoencoder-based 
approximations -- following the work pioneered in~\cite{lee2020model}; and a large number of alternative approaches for constructing a nonlinear approximation based on a nonlinear parametrization of an 
affine space~\cite{welper2017interpolation, cagniart2019model, black2020projection} -- usually through a transformation of the underlying spatial domain. Many of these PMOR methods have demonstrated, to
various degrees of success, the ability to achieve for the same level of accuracy, a lower dimension $n$ than the traditional PMOR approach based on the standard affine approximation. However,
only the PMOR method based on most-appropriate local right ROBs~\cite{amsallem2012nonlinear} was demonstrated for nonlinear,
convection-dominated problems in three dimensions, on complex (non tensor-product)
geometries, and shown to reduce wall clock time by orders of magnitude while maintaining a high-level of accuracy for spatio-temporal quantities of interest (QoI)s. Specifically, the PMOR method based 
on a piece-wise affine approximation of the solution was shown in~\cite{washabaugh2012nonlinear} to perform in real-time on a laptop, for parametric steady-state RANS computations associated with the 
NASA Common Research Model~\cite{rivers2014experimental} and an HDM of dimension greater than $N = 68\times 10^6$, in the transonic regime characterized by shocks, at high Reynolds number 
($Re = 5 \times 10^6$), and in a parameter space $\mathcal D$ of dimension $N_{\mathcal D} = 4$. Nevertheless, it is reasonable to expect this PMOR method to face increasing difficulties at
delivering a similar performance for increasing values of $N_{\mathcal D}$.

For this reason, yet another nonlinear approximation approach is developed in this paper for mitigating the Kolmogorov $n$-width barrier in the context of highly nonlinear, convection-dominated, very 
high-dimensional computational models. Following the observation noted above about the relationship between PMOR and semi-discretization, this paper proposes to construct a polynomial approximation 
$\tilde{\mathbf{u}}$ of degree $p \ge 2$, fully develops the corresponding PMOR method for $p = 2$ -- that is, the case of a quadratic approximation manifold -- and tailors the
hyperreduction method ECSW to suit this new approach to PMOR.
To this end, it is first noted that the idea of a PMOR method based on a quadratic approximation has recently been presented in~\cite{jain2017quadratic} for the reduction of a special class
of nonlinear structural dynamics models characterized by a constant mass matrix and a linear viscous damping. In this method, the construction of the quadratic approximation is {\it simulation-free}, 
in the sense that it is performed {\it a priori}, using linearization around an equilibrium position and the pre-computation of eigen modes of the linearized HDM of interest as well as their derivatives
with respect to the generalized coordinates of the approximation. Consequently, the scope of applications of this quadratic PMOR method is limited to a special class of second-order dynamical
systems, to those HDMs for which a sufficient number of eigenmodes is computable, and to those applications for which a modal analysis based on eigen vectors can deliver the desired level of accuracy. 
Hence, this scope excludes in particular convection-dominated flow problems in general and large-scale CFD-based HDMs in particular (for example, see~\cite{xie2014proper}).
On the other hand, the approach proposed in this paper for constructing a quadratic approximation $\tilde{\mathbf{u}}$ builds on the data-driven approach underlying the traditional construction of 
PROMs; is application-independent; is linearization-free and therefore is deemed to be more robust for highly nonlinear problems.
Most importantly, if for a nonlinear, convection-dominated flow problem a PMOR method equipped with the traditional affine 
approximation delivers a certain level of accuracy using a PROM of dimension $n_1$, it delivers a similar level of accuracy 
when equipped with the proposed quadratic approximation manifold using a counterpart PROM of much smaller dimension 
$n_2 \sim \sqrt n_1$.

Because the work reported in this paper is motivated by unsteady, convection-dominated, turbulent flows and given the established track records for such problems of 
LSPG~\cite{washabaugh2012nonlinear, carlberg2013gnat, grimberg2020stability} and ECSW~\cite{grimberg2021mesh}, the quadratic approximation manifold proposed in this paper is fully developed in 
the contexts of implicit time-discretization and LSPG equipped with ECSW. However, it is noted that the main contribution of this paper is equally applicable to the Galerkin framework for PMOR as well as
second-order dynamical systems, including nonlinear structural dynamics HDMs with configuration-dependent mass matrices, nonlinear damping forces, and configuration-dependent external forces/moments. 
Furthermore, because the procedure proposed in this paper for constructing a quadratic approximation manifold is independent of any parametric setting and its relative merits in terms of wall clock time 
reduction for a desired level of accuracy can be assessed independently of any such setting, this procedure is presented, described, and evaluated in this paper without reference to any parameter space $\mathcal D$ -- except time, which can always be
interpreted as the parameter of a one-dimensional (1D) parameter space $\mathcal D$.

To this end, the remainder of this paper is organized as follows. Section~\ref{sec:NLPMOR} provides a brief overview of nonlinear PMOR, incorporating a succint discussion of Galerkin and Petrov-Galerkin
projections, hyperreduction, and the Kolmogorov barrier, to keep this paper as self-contained as possible. Section~\ref{sec:QUAD} first introduces the idea of substituting the traditional affine subspace
approximation with a higher-order polynomial approximation of degree $p$, then fully develops the case $p = 2$. Specifically, this section describes in details the proposed computational procedure
for constructing the quadratic approximation manifold, outlines a heuristic for determining its appropriate dimension, 
summarizes in the form of a numerical algorithm the overall computational procedure, 
and briefly discusses the computational resources it requires. 
Section~\ref{sec:IMPACT} delineates the impact of the proposed quadratic approximation manifold on various aspects of 
LSPG and ECSW. Section~\ref{sec:APP} focuses on the RANS prediction of the convection-dominated
turbulent flow around the Ahmed body~\cite{ahmed1984some} and highlights the significant impact of the proposed quadratic approximation manifold on the performance of LSPG equipped with ECSW, in terms 
of accuracy and wall clock time reduction for a given level of accuracy. Finally. Section~\ref{sec:CONC} concludes this paper.

\section{Nonlinear projection-based model order reduction}
\label{sec:NLPMOR}

Consider the nonlinear, semi-discrete, $N$-dimensional HDM written in first-order form as

\begin{equation}
\begin{aligned}
	\mathbf{M} \mathbf{\dot{u}}\left(t\right) + \mathbf{f}\left(\mathbf{u}\left(t\right)\right) - \mathbf{g}\left(t\right) & = \mathbf{0} \\
\mathbf{u}\left(0\right) & = \mathbf{u}^0
\label{eq:hdm}
\end{aligned}
\end{equation}
where $t$ denotes time; a dot denotes a time-derivative; $\mathbf{u}(t)\in\mathbb{R}^N$ denotes the semi-discrete solution vector associated with the HDM; 
$\mathbf{M} \in \mathbb{R}^{N \times N}$ is a constant mass-like matrix; $\mathbf{f}\left(\mathbf{u}(t)\right) \in \mathbb{R}^N$ is a nonlinear vector function representing 
a semi-discrete flux or internal force vector; $\mathbf{g}(t)\in\mathbb{R}^N$ is a nonlinear vector function representing a semi-discrete external force vector, source term, or the effect of
some time-dependent boundary conditions; and $\mathbf{u}^0 \in \mathbb{R}^N$ denotes an initial condition for the semi-discrete solution vector. This HDM encompasses counterparts associated with 
nonlinear, semi-discrete, second-order dynamical systems characterized by a constant mass-like matrix but arbitrarily complex dissipative (damping) forces, which can be written as
\begin{equation}
\begin{aligned}
	\mathbf{M}_\mathbb{S} \mathbf{\ddot{v}}(t) + \mathbf{f}_\mathbb{S}^{\text{diss}}\left(\mathbf{\dot{v}}(t)\right)+ \mathbf{f}^{\text{int}}_\mathbb{S}\left(\mathbf{v}\left(t\right)\right) - \mathbf{g}_\mathbb{S}(t) & = \mathbf{0} \\
\mathbf{\dot{v}}(0) & = \mathbf{\dot{v}}^0\\
\mathbf{v}(0) & = \mathbf{v}^0
\end{aligned}
\label{eq:hdm2}
\end{equation}
Indeed, HDMs such as~\eqref{eq:hdm2} can be rewritten in the first-order form~(\ref{eq:hdm}) using the change of variables $\mathbf{u} = \left [\dot{\mathbf{v}} \quad \mathbf{v}\right]^T$, where the 
superscript $T$ designates here and throughout the remainder of this paper the transpose.

Using the transformations described in~\cite{farhat2014dimensional}, it can be shown that the HDM~(\ref{eq:hdm}) also encompasses counterparts associated with semi-discrete second-order dynamical 
systems characterized by a configuration-dependent mass-like matrix. Hence, even though all contributions of this paper are discussed in the context of the HDM~(\ref{eq:hdm}), they are 
application-independent.

\subsection{Traditional nonlinear Galerkin and Petrov-Galerkin projection-based reduced-order models}
\label{sec:CONV}

Recall that the traditional affine subspace approximation can be expressed as
\begin{equation}
\begin{aligned}
	\tilde{\mathbf{u}}(t) = \mathbf{V} \mathbf{q}(t) + \mathbf{u}_\text{ref}
\end{aligned}
\label{eq:conventional}
\end{equation}
In the context of a dynamical system, the reference solution $\mathbf{u}_{\text{ref}}$ is often chosen to represent the initial condition. In this work, the right ROB $\mathbf{V}$ is constructed as outlined
in Section~\ref{sec:INTRO}. First, the HDM is exercised -- for example, in the simulation time-interval of interest $t \in (0, T_f]$ -- to collect a set of $N_s$ solution snapshots in the snapshot matrix 
$\mathbf{S} = \left[\mathbf{u}_1, \dots, \mathbf{u}_i, \dots, \mathbf{u}_{N_s}\right]$, where the subscript $i$ designates the $i$-th solution snapshot. Then, $\mathbf{S} \in \mathbb{R}^{N\times N_s}$ 
is compressed using the thin SVD
$$ \mathbf{S} = \mathbf{U}_{\mathbf{S}} \mathbf{\Sigma}_{\mathbf{S}} \mathbf{Y}_{\mathbf{S}}^T$$
where $\mathbf{U}_{\mathbf{S}} \in \mathbb{R}^{N \times k}$ is the matrix of left singular vectors and spans the range of $\mathbf{S}$; $\mathbf{\Sigma}_{\mathbf{S}} \in \mathbb{R}^{k \times k}$ is the 
diagonal matrix of nonzero singular values $\sigma_{\mathbf{S}, i}, i \in \left\{1, \dots, k\right\}$, stored in the order $\sigma_{\mathbf{S}, 1} \geq \sigma_{\mathbf{S}, 2} \geq \dots 
\sigma_{\mathbf{S}, k} > 0$; $\mathbf{Y}_{\mathbf{S}} \in \mathbb{R}^{N_s \times k}$ is the matrix of right singular vectors; and $k \leq \min \left(N, N_s\right)$ denotes the rank of $\mathbf{S}$. 
Finally, $\mathbf{V} \in \mathbb{R}^{N \times n}$ is identified with the first $n \ll N$ columns of $\mathbf{U}_{\mathbf{S}}$. It is well known that for a given $n$, 
$\mathbf{V} \in \mathbb{R}^{N\times n}$ is optimal in the sense that it is the solution to the optimization problem

\begin{equation*}
	\arg \min_{\mathbf{X} \in \mathbb{R}^{N \times n}}\left\| \mathbf{S} - \mathbf{X} \mathbf{X}^T \mathbf{S} \right\|_F^2
\end{equation*}
where $\|\cdot\|_F$ denotes the Frobenius norm. The dimension $n$ of $\mathbf{V}$ is usually determined such that the ratio of the sum of the truncated and squared singular values, and that of all 
squared singular values, is such that
\begin{equation}
	\frac{\sum\limits_{i = 1}^n \sigma_{\mathbf{S}, i}^2}{\sum\limits_{j = 1}^k \sigma_{\mathbf{S}, j}^2} \geq 1 - \varepsilon_{\mathbf{S}}
\label{eq:singular_value_energy_criteria}
\end{equation}
where $\varepsilon_{\mathbf{S}}$ is a user-defined tolerance.

Substituting~\eqref{eq:conventional} in~\eqref{eq:hdm} leads to the nonlinear, semi-discrete, time-dependent residual equation and its associated initial condition
\begin{equation*}
\begin{aligned}
	\mathbf{r}\left(\tilde{\mathbf{u}}\right) = \mathbf{r}\left(\mathbf{V q}(t) + \mathbf{u}_{\text{ref}}, t\right) & = \mathbf{M} \mathbf{V} \mathbf{\dot{q}}(t) + \mathbf{f}\left(\mathbf{V} \mathbf{q}(t) + \mathbf{u}_{\text{ref}}\right) - \mathbf{g}\left(t\right)\\
	\mathbf{V} \mathbf{q}(0) + \mathbf{u}_{\text{ref}} & = \mathbf{u}^0
\end{aligned}
\end{equation*}
The time-discretization of the first of the above equations by a preferred {\it implicit} scheme leads at each time-step $m+1$ to a nonlinear system of algebraic equations of the form
\begin{equation} 
	\mathbf{r}^{m+1}(\tilde{\mathbf{u}}^{m+1}) = \mathbf{r}^{m+1}\left(\mathbf{V} \mathbf{q}^{m+1} + \mathbf{u}_{\text{ref}}, t^{m+1}\right)  = \mathbf{0}
	\label{eq:hdmwconv}
\end{equation}
where $\mathbf{r}^{m+1}\in \mathbb{R}^N$ and the superscript $m+1$ (or $m$) designates here and throughout the remainder of this paper a discrete quantity evaluated at time $t^{m+1}$ (or $t^m$). 
The $N$-dimensional system~\eqref{eq:hdmwconv} is overdetermined as it governs $n \ll N$ unknowns -- namely, the generalized coordinates of the subspace approximation~\eqref{eq:conventional} stored
in $\mathbf{q}^{m+1}$. Thus, the unknowns represented by the reduced-order vector $\mathbf{q}^{m+1}$ are constrained by enforcing the orthogonality of the discrete residual $\mathbf{r}^{m+1}$ to a left 
ROB $\mathbf{W} \in \mathbb{R}^{N \times n}$, which can be written as
\begin{equation}
\begin{aligned}
	\mathbf{W}^T\mathbf{r}^{m+1}\left(\mathbf{V} \mathbf{q}^{m+1} + \mathbf{u}_{\text{ref}}, t^{m+1}\right) = \mathbf{0  }
\end{aligned}
\label{eq:consresidual}
\end{equation}
If $\mathbf{W} = \mathbf{V}$ by choice, equation~\eqref{eq:consresidual} above is a discrete, nonlinear Galerkin PROM associated with the HDM~\eqref{eq:hdm}. If not, it is a Petrov-Galerkin counterpart. 
As already stated in Section~\ref{sec:INTRO}, for HDMs grounded in non-elliptic PDEs, the Petrov-Galerkin projection offers freedom in choosing $\mathbf{W}$ to achieve, for example, numerical
stability~\cite{amsallem2012stabilization}, or some notion of approximation optimality~\cite{carlberg2011efficient, carlberg2013gnat}.

For the sake of generality and because the main focus of this work is on highly nonlinear, unsteady, convection-dominated problems (which are typically non-elliptic) discretized by an implicit scheme, 
a Petrov-Galerkin projection is assumed throughout the remainder of this paper. More specifically, the LSPG framework for PMOR is adopted for the reasons outlined in Section~\ref{sec:INTRO}. In 
this case, at each time-step $m+1$ and Gauss-Newton iteration $\ell+1$, the left ROB is constructed as~\cite{carlberg2011efficient, carlberg2013gnat}
\begin{equation} 
	\mathbf{W}^{m+1, \ell+1} = \mathbf{J}^{m+1, \ell} \, \mathbf{V}
	\label{eq:WforLSPG}
\end{equation}
where
\begin{equation} 
	\mathbf{J}^{m+1, \ell} = 
			    \pdv {\mathbf{r}^{m+1}}{\tilde{\mathbf{u}}}\left(\tilde{\mathbf{u}}^{m+1, \ell}\right) = \pdv {\mathbf{r}^{m+1}}{\tilde{\mathbf{u}}}\left({\mathbf{Vq}}^{m+1, \ell} + \mathbf{u}_{\text{ref}}\right),
	\quad \mathbf{J}^{m+1, \ell} \in \mathbb{R}^{N \times N}
	\label{eq:jacobian} 
\end{equation}
and solving the discrete, nonlinear PROM equation~\eqref{eq:consresidual} becomes equivalent to solving the optimization 
problem (see~\cite{grimberg2020stability} for a mathematical proof)
\begin{equation}
\begin{aligned}
	\mathbf{q}^{m + 1} = \arg \min_{\mathbf{x} \in \mathbb{R}^n} \left\| \mathbf{r}^{m+1}\left(\mathbf{V x} + \mathbf{u}_{\text{ref}}, t^{m+1}\right)\right\|_2^2
\end{aligned}
\label{eq:gaussnewton}
\end{equation}

{\it REMARK 1.} When Galerkin projection is justified -- for example, for second-order dynamical systems or HDMs grounded in elliptic PDEs for which $\mathbf{J}^{m+1} \in \mathbb{R}^{N \times N}$ is
always symmetric positive definite -- $\mathbf{W}^{m+1} = \mathbf{W} = \mathbf{V}$, $\forall m \ge 0$. In this case, it is shown in~\cite{grimberg2020stability} that solving the discrete, nonlinear PROM 
equation~\eqref{eq:consresidual} is equivalent to solving the optimization problem
\begin{equation*}
	\mathbf{q}^{m + 1} = \arg \min_{\mathbf{x} \in \mathbb{R}^n} \left \| \left(\mathbf{J}^{m+1}\right)^{-1}\mathbf{r}^{m+1}\left(\mathbf{V x} + \mathbf{u}_{\text{ref}}, t^{m+1}\right)\right \|_{\mathbf{J}^{m+1}}
\end{equation*}

\subsection{Hyperreduction using ECSW}
\label{sec:ECSW}

For many nonlinear problems of practical interest, the complexity of the solution of problem~\eqref{eq:consresidual} -- and for this matter, that of the processing of any nonlinear PROM -- scales
not only with the small dimension $n$ of the PROM, but also with the large dimension $N$ of the HDM. Hyperreduction~\cite{ryckelynck2005priori} is one approach for eliminating this computational 
bottleneck. In this work, ECSW~\cite{farhat2014dimensional, farhat2015structure} is chosen for hyperreducing any constructed nonlinear PROM, because of its track record established 
in~\cite{farhat2014dimensional, farhat2015structure, grimberg2021mesh} for many different large-scale applications.

Let $\mathcal{E} = \left\{\mathbf{e}_i\right\}, i \in \left\{1, \dots, N_e\right\}$ denote the set of $N_e = \left |\mathcal{E}\right|$ mesh entities defining the discretization of the computational 
domain underlying the semi-discrete HDM~\eqref{eq:hdm}. For example, these entities may be elements in the case of FE modeling, or computational (primal or dual) cells in the case of finite volume (FV)
modeling. In all cases, the discrete, nonlinear PROM~\eqref{eq:consresidual} of dimension $n$ can be written as
\begin{eqnarray}
	\mathbf{r}_n^{m+1} \left(\mathbf{q}^{m+1}, t^{m+1}\right) &=& \mathbf{W}^T\mathbf{r}^{m+1}\left(\mathbf{V} \mathbf{q}^{m+1} + \mathbf{u}_{\text{ref}}, t^{m+1}\right) \nonumber \\
	&=& \sum_{e_i \in \mathcal{E}} \left(\mathbf{L}_{e_i} \mathbf{W} \right)^T \mathbf{r}_{e_i}^{m+1}\left(\mathbf{L}_{{e_i}^+} \left[\mathbf{V} \mathbf{q}^{m+1} + \mathbf{u}_\text{ref}\right], t^{m+1}\right)
\label{eq:reducedResidual}
\end{eqnarray}
where:
\begin{itemize}
	\item the notation $\mathbf{\dagger} \left [\bullet \right ]$ is used here and throughout the remainder of this paper to indicate that the quantity $\dagger$ is to be multiplied by the quantity 
		$\bullet$ and not that $\dagger$ is a function of $\bullet$.
	\item $\mathbf{r}_{e_i}^{m+1}\left(\mathbf{L}_{{e_i}^+} \left[\mathbf{V} \mathbf{q}^{m+1} + \mathbf{u}_\text{ref}\right], t^{m+1}\right) \in \mathbb{R}^{d_{e_i}}$ represents the contribution of 
		the individual mesh entity $e_i$ to the discrete residual $\mathbf{r}^{m+1}$ and $d_{e_i}$ denotes the number of dofs attached to $e_i$.  
	\item $\mathbf{L}_{e_i} \in \left\{0, 1\right\}^{d_{e_i} \times N}$ is a Boolean matrix selecting the mesh entity $e_i$ to which $d_{e_i}$ dofs are attached.  
	\item $\mathbf{L}_{{e_i}^+} \in \left\{0, 1\right\}^{d_{{e_i}^+} \times N}$ is a Boolean matrix selecting the union of the mesh entity $e_i$ and its neighbors participating in the same stencil
		of the chosen semi-discretization scheme, to which $d_{{e_i}^+}$ dofs are attached.
	\item In the case of LSPG, the left ROB $\mathbf{W}$ is given in~\eqref{eq:WforLSPG} and~\eqref{eq:jacobian}. In this case, $\mathbf{W}$ is independent of $\mathbf{q}^{m+1}$ and therefore the 
		notation of~\eqref{eq:reducedResidual} is justified even in the context of LSPG. This is because from ~\eqref{eq:WforLSPG} and~\eqref{eq:jacobian}, it follows that when the projected 
		discrete residual~\eqref{eq:reducedResidual} is expressed for $\mathbf{q}^{m+1, \ell+1}$, $\mathbf{W}$ is a function of $\mathbf{q}^{m+1, \ell}$ -- which is independent of 
		$\mathbf{q}^{m+1, \ell+1}$.
\end{itemize} 

The main idea underlying ECSW is to approximate a projected quantity such as~\eqref{eq:reducedResidual} using a cubature approach whose complexity is independent of $N$, as follows

\begin{eqnarray} 
	\mathbf{r}_n^{m+1} \left(\mathbf{q}^{m+1}, t^{m+1}\right) &=& \mathbf{W}^T\mathbf{r}^{m+1}\left(\mathbf{V} \mathbf{q}^{m+1} + \mathbf{u}_{\text{ref}}, t^{m+1}\right)  \nonumber\\
	&\approx& \sum_{e_i \in \widetilde{\mathcal{E}}} \xi_{e_i} \left(\mathbf{L}_{e_i} \mathbf{W} \right)^T \mathbf{r}_{e_i}^{m+1}\left(\mathbf{L}_{{e_i}^+} 
	\left[\mathbf{V} \mathbf{q}^{m+1} + \mathbf{u}_{\text{ref}}\right], t^{m+1}\right) 
	\label{eq:hyperreducedResidual}
\end{eqnarray}
where $\widetilde{\mathcal{E}}$ is a subset of $\mathcal{E}$ defining a ``reduced mesh'' -- that is, $n_e = \left |\widetilde{\mathcal{E}} \right| \ll N_e = \left |\mathcal{E}\right |$ -- and 
interpretable as the set of points of the cubature; and $\left \{\xi_{e_1}, \dots, \xi_{e_{n_e}}\right\}$ is the set of {\it positive}, real-valued weights of the cubature approximation.

For implicit time-discretizations, the Jacobian with respect to $\tilde{\mathbf{u}}$ of the discrete, nonlinear PROM~\eqref{eq:consresidual} of dimension $n$ can be written as
\begin{eqnarray*}
	        \mathbf{J}_n^{m+1} \left(\mathbf{q}^{m+1}, t^{m+1}\right) &=& \mathbf{W}^T\mathbf{J}^{m+1}\left(\mathbf{q}^{m+1}, t^{m+1}\right)
		= \mathbf{W}^T  \pdv {\mathbf{r}^{m+1}}{\tilde{\mathbf{u}}}\left(\mathbf{Vq}^{m+1} + \mathbf{u}_{\text{ref}}, t^{m+1}\right)\nonumber \\ 
		&=& \sum_{e_i \in \mathcal{E}} \left(\mathbf{L}_{e_i} \mathbf{W} \right)^T \mathbf{J}_{e_i}^{m+1}\left(\mathbf{L}_{{e_i}^+} \left[\mathbf{V} \mathbf{q}^{m+1} + \mathbf{u}_\text{ref}\right], t^{m+1}\right) 
		\label{eq:reducedJacobian}
\end{eqnarray*}
where $\mathbf{J}_{e_i}^{m+1}\left(\mathbf{L}_{{e_i}^+} \left[\mathbf{V} \mathbf{q}^{m+1} + \mathbf{u}_\text{ref}\right], t^{m+1}\right) \in \mathbb{R}^{d_{e_i}\times d_{e_i}}$ represents the contribution
of the individual mesh entity $e_i$ to the discrete Jacobian $\mathbf{J}^{m+1}$ and $d_{e_i}$ denotes as before the number of dofs attached to $e_i$.
Since $\displaystyle{\mathbf{W}^T \partial{\mathbf{r}^{m+1}}/\partial{\tilde{\mathbf{u}}}
= \partial{\left(\mathbf{W}^T\mathbf{r}^{m+1}\right)}/\partial{\tilde{\mathbf{u}}}}$ 
(see justification provided in the fifth bullet above), it follows that $\mathbf{J}_n^{m+1}\left(\mathbf{q}^{m+1}, t^{m+1}\right)$ is the Jacobian of the projected residual~\eqref{eq:reducedResidual}.
Therefore, $\mathbf{J}_n^{m+1}\left(\mathbf{q}^{m+1}, t^{m+1}\right)$ can be approximated using the same cubature rule as in~\eqref{eq:hyperreducedResidual}, as follows
\begin{eqnarray} 
	\mathbf{J}_n^{m+1} \left(\mathbf{q}^{m+1}, t^{m+1}\right) &=& \mathbf{W}^T\mathbf{J}^{m+1}\left(\mathbf{q}^{m+1}, t^{m+1}\right) 
	= \mathbf{W}^T  \pdv {\mathbf{r}^{m+1}}{\tilde{\mathbf{u}}}\left(\mathbf{Vq}^{m+1} + \mathbf{u}_{\text{ref}}, t^{m+1}\right)\nonumber \\
	&\approx& \sum_{e_i \in \widetilde{\mathcal{E}}} \xi_{e_i} \left(\mathbf{L}_{e_i} \mathbf{W} \right)^T \mathbf{J}_{e_i}^{m+1}\left(\mathbf{L}_{{e_i}^+} 
	\left[\mathbf{V} \mathbf{q}^{m+1} + \mathbf{u}_{\text{ref}}\right], t^{m+1}\right) 
	\label{eq:hyperreducedJacobian}
\end{eqnarray}

Specifically, ECSW computes the reduced mesh $\widetilde{\mathcal{E}}$ and its associated set of weights $\left \{\xi_{e_1}, \dots, \xi_{e_{n_e}}\right\}$ using a machine learning 
approach. Essentially, this approach trains the approximation~\eqref{eq:hyperreducedResidual} -- from which the counterpart approximation~\eqref{eq:hyperreducedJacobian} can be 
derived --  on a subset of the solution snapshots already collected in the snapshot matrix $\mathbf{S}$, as described below.

Let $\mathbf{C} = [c_{l e}] \in \mathbb{R}^{N_h \times N_e} $ and $\mathbf{d} = [d_l] \in \mathbb{R}^{N_h}$ be the matrix and vector defined as follows
\begin{equation}
\begin{aligned}
	c_{l e} & = \left(\mathbf{L}_e \mathbf{W}\right)^T \mathbf{r}_e\left(\mathbf{L}_{e^+} \left[\mathbf{V V}^T\left( \mathbf{u}_{l} - \mathbf{u}_\text{ref} \right) + 
	\mathbf{u}_\text{ref}\right], t^{l}\right), 
	& l = 1, \dots, N_h \\ 
	d_l & = \sum_{e \in \mathcal{E}} c_{l e}, & l = 1, \dots, N_h
\end{aligned}
\label{eq:training}
\end{equation}
where $N_h \le N_s$ denotes the size of the subset of the solution snapshots collected in $\mathbf{S}$ chosen for training the cubature approximation~\eqref{eq:hyperreducedResidual}.
Note that in~\eqref{eq:training} above: 
\begin{itemize}
	\item $\mathbf{V V}^T\left( \mathbf{u}_l - \mathbf{u}_{\text{ref}} \right)$ is the orthogonal projection of the collected solution snapshot $\mathbf{u}_l$ 
		onto the subspace represented by $\mathbf{V}$. This emphasizes that the ECSW training is performed for the PROM and not HDM predictions, consistently with~\eqref{eq:hyperreducedResidual}.
	\item $\mathbf{C}$ and $\mathbf{d}$ verify 
		\begin{equation} 
			\mathbf{C} \left[\boldsymbol{\xi} = \mathbf{1}\right] = \mathbf{d} 
			\label{eq:perfect}
		\end{equation}
		where $\boldsymbol{\xi} \in \mathbb{R}^{N_e}$ is the vector of weights of the cubature assuming that $\widetilde{\mathcal E} = \mathcal{E}$ and 
		$\mathbf{1} \in \mathbb{R}^{N_e}$ is the $N_e$-long vector with each entry equal to $1$. This identity is a rewriting of~\eqref{eq:reducedResidual} in matrix form.
\end{itemize}
Then, ECSW can be simply described as relaxing~\eqref{eq:perfect} and defining instead $\widetilde{\mathcal E}$ and $\boldsymbol{\xi}$ as the solution
of the optimization problem
\begin{equation}
	\min_{\boldsymbol{\xi} \in \mathbb{R}^{N_e}_{\geq 0}} \|\boldsymbol{\xi}\|_{\text{ref}} \quad s.t. \quad \|\mathbf{C}\mathbf{\xi} - \mathbf{d}\|_2 \leq \tau\|\mathbf{d}\|_2
	\label{eq:hard}
\end{equation}
where $\|\bullet\|_0$ designates the $\ell_0$-norm of $\bullet$ and $0 < \tau < 1$ is a specified training tolerance. Unfortunately, the above minimization problem is combinatorially hard. For this 
reason, several alternative formulations of~\eqref{eq:hard} were considered in~\cite{chapman2017accelerated} and the following nonnegative least squares approach 
\begin{equation}
\begin{aligned}
	\boldsymbol{\xi} = \arg \min_{\boldsymbol{\zeta} \in \mathbb{R}_{\geq 0}^{N_e}} \left\| \mathbf{C} \boldsymbol{\zeta} - \mathbf{d} \right\|_2^2
\label{eq:nnls_ecsw}
\end{aligned}
\end{equation}
equipped with the early termination criterion
\begin{equation}
	\left \| \mathbf{C} \boldsymbol{\xi} - \mathbf{d} \right\| \leq \tau \left\| \mathbf{d} \right\|_2
\label{eq:nnls_ecsw_early_termination}
\end{equation}
was found to be most practical and computationally efficient. Due to~\eqref{eq:nnls_ecsw_early_termination}, the solution of the optimization problem~\eqref{eq:nnls_ecsw} is a sparse vector 
$\boldsymbol{\xi}$ characterized by a relatively small number of nonzero entries corresponding to the mesh entities defining the reduced mesh $\widetilde{\mathcal{E}}$. 

Finally, after ECSW has computed the reduced mesh $\widetilde{\mathcal E} \subset \mathcal{E}$, the augmented counterpart $\widetilde{\mathcal{E}}^+$ is obtained by simply adding to 
$\widetilde{\mathcal{E}}$ the neighbors to its mesh entities that participate in the same stencil of the chosen semi-discretization scheme and have not already been sampled in $\widetilde{\mathcal{E}}$.

\subsection{Kolmogorov $n$-width barrier to reducibility}
\label{sec:KB}

The Kolmogorov $n$-width of a subspace $\mathcal{M}$ of dimension $n$ is defined as
\begin{equation}
d_n\left(\mathcal{M}\right) = \inf_{\substack{\mathcal{Y}_n \subseteq \mathcal{Y} \\ \dim\left(\mathcal{Y}_n\right) \leq n}} \sup_{y \in \mathcal{M}} \inf_{y_n \in \mathcal{Y}_n} \left|\left|y - y_n\right|\right|_\mathcal{Y}
\label{eq:kolmogorov_n_width}
\end{equation}
where $\mathcal{Y}$ is a linear subspace~\cite{unger2019Kolmogorov}. Effectively, the combination of the supremum and second infimum determines the maximum distance in a $\mathcal{Y}$-norm between the 
lines $y$ and $y_n$ when projecting points in $\mathcal{M}$ onto an $n$-dimensional subspace of $\mathcal{Y}$. It is a useful concept for determining the extent to which, for a fixed size $n$ of the ROB,
the traditional PROM approximation~\eqref{eq:conventional} can accurately represent the solution associated with a given HDM. It is equally useful concept for predicting computational efficiency, as for a
hyperreduced nonlinear PROM (HPROM) whose underlying HDM is grounded in a PDE, the online solution time scales primarily with the dimension $n$ of the PROM as well as with the size of the reduced mesh.

In the context of a PROM or HPROM whose underlying HDM is grounded in a PDE, the rate of convergence of the Kolmogorov $n$-width depends on the mathematical type of the PDE. The most optimistic case corresponds to a subset 
of linear coercive problems: in this case, $d_n(\mathcal{M}) \leq C e^{-\beta n}$, where $0 < C < \infty$ and $\beta > 0$ \cite{buffa2012priori}. For such problems, very low-dimensional 
PROMs can deliver an acceptable accuracy. For an HDM grounded in a nonlinear hyperbolic PDE, the decay of the $n$-width of an associated PROM may be much slower: in some cases,
$d_n(\mathcal{M}) \leq 1/4\left(n^{-1/2}\right)$~\cite{greif2019decay}, which explains why for such HDMs the Kolmogorov $n$-width is often referred to in the 
literature as a barrier to reducibility. For example, for a discrete HDM of dimension greater than $N = 17\times 10^6$ of the convection-dominated turbulent flow around the Ahmed 
body~\cite{ahmed1984some}, which is a popular benchmark CFD problem in the automotive industry, an unsteady LSPG PROM -- and therefore a PROM parameterized by time only -- based on the traditional global
subspace approximation~\eqref{eq:conventional} requires a dimension $n \approx 600$ to achieve sufficient accuracy~\cite{grimberg2021mesh}. On the other hand, an LSGP PROM based on a 100-subdomain 
partition of the solution manifold and the piece-wise affine approximation approach proposed in~\cite{amsallem2012nonlinear} achieves the same level of accuracy using an average dimension 
$n \approx 11$. However, as explained in Section~\ref{sec:INTRO}, in the context of a high-dimensional parameter space $\mathcal{D}$, an LSPG PROM based on the same piece-wise affine approximation can 
be expected to require a much larger average dimension $n$ and potentially be unafforable. For this reason, for nonlinear, convection-dominated, first-order hyperbolic problems, there is a significant 
interest in developing alternative, more efficient nonlinear approximation manifolds for PMOR.

\section{Data-driven quadratic approximation manifold}
\label{sec:QUAD}

The proposed data-driven, higher-order polynomial approximation manifold is a compromise between: on one extreme, the comprehensiveness of an arbitrarily nonlinear approximation manifold such as that 
based on a convolutional autoencoder~\cite{lee2020model} and its potential for accuracy; and on another extreme, the computational efficiency of the training step of the simulation-free quadratic 
manifold approximation proposed for structural dynamics applications in~\cite{jain2017quadratic}. The {\it total} offline cost of its quadratic instance -- defined here as the computational cost 
associated with the construction/training of both the matrix coefficients of the quadratic solution approximation and the reduced mesh needed for hyperreduction -- is similar to if not less than that 
of the POD method based on the traditional affine approximation~\eqref{eq:conventional}. However, for a fixed dimension $n$ of the target nonlinear PROM, its accuracy is far superior. Furthermore, the 
proposed higher-order polynomial approximation manifold approach is amenable to a piece-wise implementation similar to that developed in~\cite{amsallem2012nonlinear}, but such an implementation is work 
in progress.

\subsection{Higher-order polynomial approximation manifolds}

Again, PMOR is essentially a semi-discretization method for PDEs based on global shape functions; it can be interpreted as a Ritz approximation method. Therefore, a {\it natural} approach for improving 
its performance for problems where the affine approximation~\eqref{eq:conventional}, which is a polynomial approximation of degree $p=1$, is inefficient is to consider a higher-order polynomial 
approximation of degree $p \geq 2$. Such an approximation can be written as
\begin{equation}
\begin{aligned}
	\tilde{\mathbf{u}}(t) & = \sum_{i = 1}^{p} \mathbf{G}_{p-i+1} \left[\mathbf{q}^{\otimes p-i+1}(t)\right] \, + \mathbf{u}_{\text{ref}} 
\end{aligned}
\label{eq:polynomial}
\end{equation}
where $\mathbf{G}_j \in \mathbb{R}^{N \times n^{j}}$, the superscript $\otimes j$ designates the $j$-fold Kronecker product, and therefore $\mathbf{q}^{\otimes j} \in \mathbb{R}^{n^j}$. Observe that if
$\mathbf{G}_1$ is chosen to be the matrix $\mathbf{V}\in \mathbb{R}^{N \times n}$ that was previously referred to as the right ROB, the traditional affine approximation~\eqref{eq:conventional} becomes 
the particular case of~\eqref{eq:polynomial} corresponding to $p=1$.

The present work focuses on the case $p = 2$ -- that is, the solution approximation 
$\tilde{\mathbf{u}}(t) = \mathbf{G}_2\left[\mathbf{q}^{\otimes 2}(t)\right] + \mathbf{G}_1 \mathbf{q}(t) + \mathbf{u}_{\text{ref}}$. Following the aforementioned observation, $\mathbf{G}_1$ is set by 
design to $\mathbf{G}_1 = \mathbf{V}\in \mathbb{R}^{N \times n}$ and $\mathbf{V}$ is constructed using the same data-driven
procedure -- but not necessarily the same dimension -- as in the case of the traditional affine approximation~\eqref{eq:conventional}. 
In this case, the approximation~\eqref{eq:polynomial} with $p=2$ can be rewritten using a notation closer to that 
of~\eqref{eq:conventional} as follows
\begin{equation}
	\tilde{\mathbf{u}}(t) = \mathbf{H} \left[\mathbf{q}^{\otimes 2}(t)\right] + \mathbf{V q}(t) + \mathbf{u}_{\text{ref}} = 
	\mathbf{H} \left[\mathbf{q}(t) \otimes \mathbf{q}(t)\right] + \mathbf{V q}(t) + \mathbf{u}_{\text{ref}}
\label{eq:quad_approx}
\end{equation}
where $\mathbf{H} \in \mathbb{R}^{N \times n^2}$ and $\otimes$ represents the vectorized Kronecker product such that $\mathbf{q} \otimes \mathbf{q} \in \mathbb{R}^{n^2}$. 

It is noted here that an approximation of the form given in~\eqref{eq:quad_approx} has been recently used in~\cite{mcquarrie2021data} in a different context, namely, that of constructing a 
{\it nonintrusive}, {\it projection-free}, reduced-order model based on the notion of operator inference. Hence, both this work and the work presented earlier in~\cite{mcquarrie2021data} share the 
task of constructing numerically a coefficient matrix $\mathbf{H}$ governing the quadratic term of a quadratic approximation. However, even with regards to this specific task, the global contexts, the 
numerical approaches, and most importantly, the exploitations of approximations of the form given in~\eqref{eq:quad_approx} for constructing surrogate models, and the types and performances of the 
surrogate models discussed here and in~\cite{mcquarrie2021data} are vastly different.

\subsection{Two-step construction of a quadratic approximation manifold}
\label{sec:QM}

Here, a two-step approach is described for constructing the matrix $\mathbf{H}$ of the quadratic approximation~\eqref{eq:quad_approx}.

In the first step, the reference solution $\mathbf{u}_{\text{ref}} \in \mathbb{R}^N$ is chosen as usual and the matrix $\mathbf{V} \in \mathbb{R}^{N\times n}$, where $n \ll N$, is constructed as in the 
method of snapshots for POD~\cite{sirovich1987turbulence} using however SVD for data compression -- and as can be expected, a lower dimension $n$ for the same level of accuracy.

In the second step, $\mathbf{H} \in \mathbb{R}^{N \times n^2}$ is determined such that for all solution snapshots collected in $\mathbf{S} \in \mathbb{R}^{N \times N_s}$, the errors of the quadratic 
approximation~\eqref{eq:quad_approx} are minimized.

The two-step aspect of the approach outlined above for determining the matrix coefficients of the quadratic solution approximation~\eqref{eq:quad_approx} offers the following advantages over an
alternative approach where both $\mathbf{V} \in \mathbb{R}^{N\times n}$ and $\mathbf{H} \in \mathbb{R}^{N \times n^2}$ are simultaneously determined: 
\begin{itemize}
	\item Full column rank and orthogonality of $\mathbf{V}$ -- for example, $\mathbf{V}^T \mathbf{V} = \mathbf{I}$ -- which have been very convenient in the numerical 
		implementation of PMOR equipped with the traditional subspace approximation~\eqref{eq:conventional}, at no additional computational effort. 
	\item Lower computational complexity due to the divide-and-conquer nature of the approach where: first, $\mathbf{V}$ is computed; and then, $\mathbf{H}$ is determined given $\mathbf{V}$.
\end{itemize}

Whether $\mathbf{V}$ and $\mathbf{H}$ are determined as outlined above or simultaneously, the resulting construction of the quadratic approximation manifold underlying~\eqref{eq:quad_approx}
is data-driven. It is agnostic to the specific form of the nonlinear HDM and therefore applicable to the PMOR of any nonlinear HDM. Furthermore, as the structure of the nonlinear approximation term is 
pre-determined, $\mathbf{H} \in \mathbb{R}^{N \times n^2}$ can be computed row-by-row. Specifically, it is shown in Section~\ref{sec:CQUAD} that each row of $\mathbf{H}$ can be determined by 
solving an independent linear least squares problem. This way, $\mathbf{H}$ can be efficiently determined using furthermore embarrassingly parallel computations.

\subsection{Computation of the coefficient matrix of the quadratic term of the approximation}
\label{sec:CQUAD}

For each solution snapshot $\mathbf{u}_l$ collected in $\mathbf{S} \in \mathbb{R}^{N \times N_s}$, let $\mathbf{e}_l \in \mathbb{R}^{N}$ denote the error associated with its orthogonal
projection on the subspace represented by $\mathbf{V}$ -- that is,
\begin{equation} 
	\mathbf{e}_l = \mathbf{u}_l - \mathbf{V}\mathbf{q}_l - \mathbf{u}_{\text{ref}}, \  l = 1, \dots, N_s
	\label{eq:ei}
\end{equation}
If the vector of generalized coordinates $\mathbf{q}_l$ is computed such that the corresponding error $\mathbf{e}_l$ is orthogonal to the subspace represented by $\mathbf{V}$ and if by 
construction $\mathbf{V}$ is orthogonal, it follows from~\eqref{eq:ei} that
\begin{equation} 
	\mathbf{q}_l = \mathbf{V}^T \left(\mathbf{u}_l - \mathbf{u}_{\text{ref}}\right), \ l = 1, \dots, N_s
	\label{eq:qi}
\end{equation}

Assembling all approximation errors~\eqref{eq:ei} in the matrix $\mathbf{E}\in \mathbb{R}^{N\times N_s}$ and all vectors of generalized coordinates~\eqref{eq:qi} in 
$\mathbf{Q} \in \mathbf{R}^{n^2\times N_s}$ as follows
\begin{equation}
\begin{aligned}
\mathbf{E} & = \left[\mathbf{e}_1 \, \dots \, \mathbf{e}_{N_s} \right]\\
\mathbf{Q} & = \left[ \mathbf{q}_1 \otimes \mathbf{q}_1 \, \dots \, \mathbf{q}_{N_s} \otimes \mathbf{q}_{N_s}\right]\\
\end{aligned}
\label{eq:EandQ}
\end{equation}
the second step of the two-step approach outlined in Section~\ref{sec:QM} for constructing the matrix $\mathbf{H} \in \mathbb{R}^{N \times n^2}$ can be written as
\begin{equation}
	\mathbf{H} = \arg \min_{\mathbf{H}^{\prime} \in \mathbb{R}^{N \times n^2}} \left|\left| \mathbf{E} - \mathbf{H}^{\prime} \mathbf{Q} \right|\right|_F
\label{eq:H_opt}
\end{equation}

Let $[\mathbf{E}]_{i,1}$ and $[\mathbf{H}]_{i,1}$, $i = 1, \dots, N$, denote the row partitioning of each of $\mathbf{E}$ and $\mathbf{H}$, respectively. Using this partitioning, the minimization
problem~\eqref{eq:H_opt} can be rewritten as
\begin{equation} 
	\mathbf{H} = \arg \min_{\mathbf{H}^{\prime} \in \mathbb{R}^{N \times n^2}} \sum\limits_{i=1}^{N}\left|\left| [\mathbf{E}]_{i,1} - [\mathbf{H}^{\prime}]_{i,1} \mathbf{Q} \right|\right|_2^2
	\label{eq:H_opt_row} 
\end{equation}
From~\eqref{eq:H_opt} and~\eqref{eq:H_opt_row}, it follows that $\mathbf{H}$ can be computed row-by-row, by solving the following sequence of $N$ independent minimization problems
\begin{equation} 
	\mathbf{h}_i = [\mathbf{H}]_{i,1} = \arg \min_{{\mathbf{h}^{\prime}_i} \in \mathbb{R}^{1 \times n^2}} \left|\left| [\mathbf{E}]_{i,1} - [\mathbf{h}^{\prime}]_i \mathbf{Q} \right|\right|_2^2, \quad i = 1, \dots, N 
	\label{eq:H_opt_rowwise}
\end{equation}
This distributed approach for computing the matrix $\mathbf{H}$ is preferred over its counterpart outlined in~\eqref{eq:H_opt} because it is less computationally intensive -- it incurs a sequence of
smaller minimization problems; and furthermmore, it can be implemented on a massively parallel computing system in an embarrassingly parallel manner.

Unfortunately, all optimization problems implied in~\eqref{eq:H_opt_row} and~\eqref{eq:H_opt_rowwise} are susceptible to over-fitting. For this reason, the solution approach~\eqref{eq:H_opt_rowwise}
is modified to include a Tikhonov regularization~\cite{hansen1998rank} using the classical Tikhonov matrix $\mathbf{\Gamma} = \alpha \mathbf{I}$, where $\alpha > 0$ and $\mathbf{I}$ denotes the 
identity matrix. This leads to
\begin{equation}
	\mathbf{h}_i = [\mathbf{H}]_{i,1} = \arg \min_{{\mathbf{h}^{\prime}_i} \in \mathbb{R}^{1 \times n^2}} \left|\left| [\mathbf{E}]_{i,1} - [\mathbf{h}^{\prime}]_i \mathbf{Q} \right|\right|_2^2 + \alpha \left|\left| [\mathbf{h}^{\prime}]_i \right|\right|_2^2, \quad i = 1, \dots, N
	\label{eq:H_opt_1row}
\end{equation}
Note that by choice, the same regularization parameter $\alpha$ is used in~\eqref{eq:H_opt_1row} for all instances $i$ = $1, \dots, N$.

Depending on the combination of the number of collected solution snapshots $N_s$ and the target dimension of the PROM $n$, each regularized
least squares problem in~\eqref{eq:H_opt_1row} can be underdetermined, 
square, or overdetermined. It is also ill-conditioned, at least for the following reason. Each column block $\mathbf{q}_j^{\otimes} = \mathbf{q}_j \otimes \mathbf{q}_j$ of $\mathbf{Q}$~\eqref{eq:EandQ} 
contains $n^2$ entries: however, due to symmetry, $n(n-1)/2$ of these are repeated entries. This specific issue can be addressed by simply excluding all duplicate entries from the construction
of $\mathbf{Q}$ to obtain in principle a redundancy-free matrix 
$\overline{\mathbf{Q}} \in \mathbb{R}^{n(n+1)/2 \times N_s}$. In this case, the corresponding entries in each row vector
$\mathbf{h}_i$ are also excluded from its definition to obtain a new row vector $\bar{\mathbf{h}}_i \in \mathbb{R}^{1 \times n(n+1)/2}$. 

In general, it is preferrable that $N_s$ and $n$ verify $N_s > n(n+1)/2$, so that each regularized least squares problem 
in~\eqref{eq:H_opt_1row} is overdetermined and can be solved using the non-truncated thin SVD of $\overline{\mathbf{Q}}$ as follows
\begin{equation}
	\overline{\mathbf{Q}} = \mathbf{U}_{\overline{\mathbf{Q}}} \mathbf{\Sigma}_{\overline{\mathbf{Q}}} \mathbf{Y}^T_{\overline{\mathbf{Q}}} \qquad \Rightarrow \qquad
	\bar{\mathbf{h}}_i^T  = \sum_{l=1}^{n_{\overline{\mathbf{Q}}}} \left(\frac{\sigma_{\overline{\mathbf{Q}}, l}^2}{\sigma_{\overline{\mathbf{Q}}, l}^2 + \alpha^2}\right) \frac{\mathbf{y}_{\overline{\mathbf{Q}}, l}^T \left[\mathbf{E}\right]_{i,1}^T}{\sigma_{\overline{\mathbf{Q}}, l}} \mathbf{u}_{\overline{\mathbf{Q}}, l}, \quad i = 1, \dots, N
	\label{eq:sol1row}
\end{equation}
where $n_{\overline{\mathbf{Q}}}$ is the dimension of $\mathbf{\Sigma}_{\overline{\mathbf{Q}}}$ -- that is, the number of singular values $\sigma_{\overline{\mathbf{Q}}, l}$ of $\overline{\mathbf{Q}}$,
$l \in \{1, \dots, n_{\overline{\mathbf{Q}}}\}$ -- and $\mathbf{u}_{\overline{\mathbf{Q}}, l}$ and $\mathbf{y}_{\overline{\mathbf{Q}}, l}$ are the $l$-th column vectors of 
$\mathbf{U}_{\overline{\mathbf{Q}}} \in \mathbb{R}^{n(n+1)/2\times n_{\overline{\mathbf{Q}}}}$ and $\mathbf{Y}_{\overline{\mathbf{Q}}} \in \mathbb{R}^{N_s \times n_{\overline{\mathbf{Q}}}}$, 
respectively~\cite{hansen1998rank}.

The computation of the row matrices $\bar{\mathbf{h}}_i^T$, $i = 1, \dots, N$~\eqref{eq:sol1row}, depends on the choice of the parameter $\alpha$ determining the strength of the regularization.
A reasonable value of this parameter can be found using the generalized cross-validation (GCV) method~\cite{golub1979generalized}, as typically done for a Tikhonov regularization. In this work,
$\alpha$ is specifically determined as follows. First, a real-valued parameter $0 < \omega \leq 1$ is introduced to specify the length $n_{\text{smp}} = \lceil\omega n_{\overline{\mathbf{Q}}}\rceil$ of 
the vector $\boldsymbol{\alpha}_{\text{smp}} = \{\alpha_1, \dots, \alpha_{n_{\text{smp}}}\}$ of trial values of $\alpha$
. This vector consists of $n_\text{smp}$ 
evenly spaced (in a logarithmic sense) entries in the interval $\left[\sigma_{\overline{\mathbf{Q}}, n_{\overline{\mathbf{Q}}}}, \sigma_{\overline{\mathbf{Q}}, 1}\right]$ of singular values of
$\overline{\mathbf{Q}}$ stored in sorted order, from the smallest to the largest. Next, for each $i$-th degree of freedom (dof) of the HDM, $i = 1, \dots, N$, GCV is applied to compute a temporary value 
$\alpha_i$ of the regularization parameter. Then, $\alpha$ is set to the mode of the sample -- that is, the value $\alpha^{\star}$ that occurs most 
often in the collection. While this approach may occasionally overregularize the least squares problems~\eqref{eq:H_opt_1row}, it is computationally efficient: in particular, it requires performing the 
thin SVD of $\overline{\mathbf{Q}}$ only once.

\subsection{Heuristic for an appropriate dimension of the quadratic approximation manifold}
\label{sec:HEURISTIC}

Here and throughout the remainder of this paper, a PROM equipped with the traditional affine approximation~\eqref{eq:conventional} is referred to as a {\it traditional} PROM, while a counterpart 
equipped with the quadratic approximation~\eqref{eq:quad_approx} is referred to as a {\it quadratic} PROM. Strictly speaking, this terminology is independent of the linearity or nonlinearity of the HDM 
and its associated PROM. In other words, a nonlinear PROM associated with a nonlinear HDM can be traditional, if the underlying solution approximation is affine or linear; or quadratic, if the 
underlying solution approximation is quadratic (and therefore nonlinear).

Recall and/or observe that:
\begin{itemize}[topsep=0pt]
	\itemsep=0pt
	\item{I1.} For problems~\eqref{eq:H_opt_1row} to be overdetermined, the number $N_s$ of solution snapshots collected in
		$\mathbf{S}\in \mathbb{R}^{N\times N_s}$ and the dimension $n$ of the target quadratic PROM must satisfy 
		$N_s > n(n+1)/2$. For parametric problems, $N_s$ is an outcome of a parameter sampling procedure 
		-- preferrably, an adaptive one such as a greedy procedure based on an errror indicator or 
		estimator (for example, see~\cite{paul2015adaptive}). In the particular case where only time is a parameter 
		-- say an unsteady turbulent flow or a nonlinear, structural dynamic computation -- $N_s$ is often simply 
		determined by a non-adaptive sampling frequency.
	\item{I2.} In the two-step approach for constructing the quadratic approximation manifold 
		underlying~\eqref{eq:quad_approx}, 
	       $\mathbf{V} \in \mathbb{R}^{N\times n}$ is constructed as in the case of the traditional subspace 
		approximation~\eqref{eq:conventional}; and therefore $n$ is typically determined using the singular value energy 
		criterion~\eqref{eq:singular_value_energy_criteria} that depends on the tolerance $\varepsilon_{\mathbf{S}}$~\eqref{eq:singular_value_energy_criteria}.
	\item{I3.} For the same reference solution $\mathbf{u}_{\text{ref}} \in \mathbb{R}^{N}$, the traditional subspace 
		approximation~\eqref{eq:conventional} depends on $Nn$ control variables that define the right ROB
		$\mathbf{V} \in \mathbb{R}^{N\times n}$ and the approximation subspace it represents; but the quadratic 
		approximation~\eqref{eq:quad_approx} depends on $Nn(n+1)/2 + Nn$ control variables that define the rows
		$\bar{\mathbf{h}}_i \in \mathbb{R}^{1 \times n(n+1)/2}$, $i = 1, \dots, N$, and $\mathbf{V} \in \mathbb{R}^{N\times n}$. 
		Based on matching the numbers of control variables for both cases, it follows that for a given nonlinear, 
		convection-dominated problem, one can reasonably expect a quadratic PROM of dimension 
		$\left(\sqrt{9+8n} -3\right)/2 < n$ (and $\left(\sqrt{9+8n}-3\right)/2 \ll n$, if $n \geq 100$) to deliver the 
		same accuracy as that of a traditional counterpart of dimension $n$. 
	\item {I4.} The solution~\eqref{eq:sol1row} of problems~\eqref{eq:H_opt_1row} is vulnerable to overregularization, which may lead to a loss of some of the capacity of the $Nn(n+1)/2$ control 
		variables defining $\bar{\mathbf{h}}_i \in \mathbb{R}^{1 \times n(n+1)/2}$, $i = 1, \dots, N$, to construct the best possible quadratic approximation in the sense defined in~\eqref{eq:H_opt}.
		This in turn can be interpreted as working effectively with a number $Nn^{\prime}(n^{\prime}+1)/2 < Nn(n+1)/2$ of these control variables -- or equivalently, with 
		a dimension $n^{\prime} < n$.
\end{itemize}

Based on the above reminders and observations, the following heuristic is proposed for selecting an appropriate dimension $n$
for a quadratic PROM:
\begin{itemize}[topsep=0pt]
	\itemsep=0pt
	\item From item I2 above,
		set $\varepsilon_{\mathbf{S}}$ in~\eqref{eq:singular_value_energy_criteria} to its usual appropriate value for constructing a right ROB $\mathbf{V}$ for a traditional PROM 
		-- say $\varepsilon_{\mathbf{S}} = 10^{-4}$ and therefore $1 - \varepsilon_{\mathbf{S}} = 99.99\%$ -- and compute the corresponding ROB dimension $n_\text{tra}$.
	\item From item I3 above, compute $n^{\prime}_\text{qua} = \left(\sqrt{9+8n_\text{tra}} - 3\right)/2$, which assumes no overregularization of the least squares problems~\eqref{eq:H_opt_1row}.
	\item From item I4 above, set $n_{\text{qua}} = (1 + \zeta) n^{\prime}_\text{qua}$, where $0 < \zeta < 0.2$, which attempts to correct for any overregularization of the least squares 
		problems~\eqref{eq:H_opt_1row} due to the choice $\alpha = \alpha^\star$.
	\item Finally, to satisfy the constraint recalled in item I1 above, set
		\begin{equation}
			n = \min \left(n_\text{qua}, \left(\sqrt{1+8N_s} - 1\right)/2 \right)
			\label{eq:CONS}
		\end{equation}
\end{itemize}

The performance of the above heuristic depends on the parameter $\zeta$. In general, the optimal value of this parameter depends on the method chosen for determining the regularization parameter
$\alpha$ -- in this case, GCV -- and its specific implementation -- in this case, through $\alpha^\star$. Hence, the heuristic described above is not meant to be the final word on how to determine the 
dimension $n$ of a quadratic PROM, but to provide an initial idea for how to approach this task. Its soundness is confirmed by the numerical results reported in Section~\ref{sec:APP} for the prediction 
of the highly nonlinear, convection-dominated, turbulent flow around and in the wake of the Ahmed body.   

\subsection{Numerical algorithm and required computational resources}
\label{sec:NACO}

The overall computational procedure proposed in this paper for constructing a quadratic approximation manifold for the
purpose of nonlinear PMOR is summarized in {\bf Algorithm 1} below.

\begin{algorithm}
	\caption{Computation of the coefficient matrices of the quadratic approximation manifold~\eqref{eq:quad_approx}.}
      \label{alg:calc_H}
      \begin{algorithmic}
	      \INPUT solution snapshot matrix $\mathbf{S} \in \mathbb{R}^{N\times N_s}$; tolerance $\varepsilon_{\mathbf{S}}$; correction factor $\zeta$; regularization parameter $\alpha^\star$ or parameter $\omega$
	      \OUTPUT dimension $n$; matrix $\mathbf{V}\in\mathbb{R}^{N \times n}$; matrix $\overline{\mathbf{H}} \in \mathbb{R}^{N \times n(n+1)/2}$
	      
	      \State $\mathbf{U}_{\mathbf{S}} \mathbf{\Sigma}_{\mathbf{S}} \mathbf{Y}_{\mathbf{S}}^T \gets \mathbf{S}$ \Comment{Compute SVD}
	      \State $\mathbf{\sigma}_{\mathbf{S}} \gets \text{diag}\left(\mathbf{\Sigma}_{\mathbf{S}}\right)$ \Comment{$\mathbf{\sigma}_{\mathbf{S}} \in \mathbb{R}^k$, $k \leq \min \left(N, N_s\right)$}
	      \State $n_\text{tra} \gets $ smallest $n$ satisfying $\left(1  - \sum\limits_{i = 1}^{n} \sigma_{{\mathbf S},i} \Big/ \sum\limits_{j = 1}^{k} \sigma_{\mathbf{S}, j}\right) \leq \varepsilon_{\mathbf{S}}$
	      \State $\mathbf{V} \gets \left[\mathbf{u}_1, \dots, \mathbf{u}_{n_\text{tra}}\right]$ \Comment{$\mathbf{u}_i$ is the $i$-th column of $\mathbf{U}_{\mathbf{S}}$}	      
	      
        \State $n_\text{qua}^\prime \gets \left(\sqrt{9 + 8 n_\text{tra}} - 3\right) / 2$
        \State $n_\text{qua} \gets (1 + \zeta) n_\text{qua}^\prime$ \Comment{See I4 in Section~\ref{sec:HEURISTIC}} for justification
        \State $n \gets \min \left(n_\text{qua}, \left(\sqrt{1+8N_s} - 1\right)/2 \right)$
        \State Truncate $\mathbf{V}$ such that $\mathbf{V} \in \mathbb{R}^{N \times n}$
	\State Declare $\mathbf{E} \in \mathbb{R}^{N \times N_s}, \overline{\mathbf{Q}} \in \mathbb{R}^{n(n+1)/2 \times N_s}, \overline{\mathbf{H}} \in \mathbb{R}^{N \times n(n+1)/2 }$
        \For{$i \gets 1$ to $N_s$}
        	\State $\mathbf{q}_i = \mathbf{V}^T\mathbf{S}[:,i]$
		\State $\mathbf{E}[:,i] \gets \mathbf{S}[:,i] - \mathbf{V} \mathbf{q}_i - \mathbf{u}_{\text{ref}}$
        	\State \parbox[t]{\dimexpr\textwidth-\leftmargin-\labelsep-\labelwidth}{$\overline{\mathbf{Q}}[:,i] \gets \texttt{unique}\left(\mathbf{q}_i \otimes \mathbf{q}_i\right)$ \Comment{\texttt{unique} removes duplicate values from the Kronecker product due to symmetry with respect to exchange of arguments}}
        \EndFor
        \State Declare $\mathbf{U}_{\overline{\mathbf{Q}}} \in \mathbb{R}^{n(n+1)/2 \times n_{\overline{\mathbf{Q}}}}, \Sigma_{\overline{\mathbf{Q}}} \in \mathbb{R}^{n_{\overline{\mathbf{Q}}} \times n_{\overline{\mathbf{Q}}}}, \mathbf{Y}_{\overline{\mathbf{Q}}} \in \mathbb{R}^{N_s \times n_{\overline{\mathbf{Q}}}} \gets \texttt{SVD}\left(\overline{\mathbf{Q}}\right)$ \Comment{Compute thin SVD}
	      \If{$\alpha^\star$ is not specified}
        	\State Declare $\mathbf{\alpha}^\text{best} \in \mathbb{R}^{N}$
        	\State Declare $n_\text{smp} \gets \lceil \omega n_{\overline{\mathbf{Q}}}\rceil$
		\State Declare $\boldsymbol{\alpha}_{\text{smp}} \in \mathbb{R}^{n_\text{smp}} \gets$ $n_\text{smp}$ samples uniformly distributed in log scale, from the maximum to the minimum singular value of $\overline{\mathbf{Q}}$
		\For{$i \gets 1$ to $N$}
			\State Declare $\mathbf{G} \in \mathbb{R}^{n_\text{smp}}$
			\For{$k \gets 1$ to $n_{\text{smp}}$}
				\State $\mathbf{G}[k] \gets \texttt{GCV}\left(\alpha[k] \right)$ \Comment{See chapter 7 of \cite{hansen1998rank} for details regarding the function \text{GCV}}
        		\EndFor
        		\State $\alpha^{\text{best}}[i] \gets \alpha$ that corresponds to the minimum of the entries of $\mathbf{G}$
        	\EndFor
        	\State $\alpha^\star$ $ \gets $ most frequently selected value of the regularization parameter stored in $\alpha^{\text{best}}$
        \EndIf
        
        \For{$i \gets 1$ to $N$}
		\State Declare $\overline{\mathbf{h}}_i \gets \mathbf{0}$
		\For{$l \gets 1 $ to $n_{\overline{\mathbf{Q}}}$}
		\State $\overline{\mathbf{h}}_i \ \texttt{+=} \ \left(\frac{\sigma_{\overline{\mathbf{Q}}, l}^2}{\sigma_{\overline{\mathbf{Q}}, l}^2 + \alpha^{{\star}^2}}\right) \frac{\mathbf{y}_{\overline{\mathbf{Q}}, l}^T \mathbf{E}[l,:]^T}{\sigma_{\overline{\mathbf{Q}}, l}} \mathbf{u}_{\overline{\mathbf{Q}}, l}$
		\EndFor
		\State $\overline{\mathbf{H}}[i,:] \gets \overline{\mathbf{h}}_i$
        \EndFor
        
      \end{algorithmic}
\end{algorithm}

\clearpage

The computational cost of {\bf Algorithm 1} has two main components: the computational cost associated with the
construction of the matrix $\mathbf{V} \in \mathbb{R}^{N\times n}$ -- which is essentially the same as that associated with the
construction of a right ROB for a traditional PROM (C1); and that associated with the construction of the row matrices
$\bar{\mathbf{h}}_i \in \mathbb{R}^{1 \times n(n+1)/2}$, $i = 1, \dots, N$ -- which is specific to the construction
of the proposed quadratic approximation manifold associated with a quadratic PROM (C2). The general conclusion, 
which is explained below, is that in the best case, C2 is a small fraction of C1 and therefore the total cost of the proposed 
approach for constructing a quadratic approximation manifold is comparable to that associated with the construction of a 
traditional subspace approximation; and in the worst case, C2 is comparable to C1 and therefore the total cost of the proposed
approach for constructing a quadratic approximation manifold is about twice that associated with the construction of a
traditional subspace approximation.

As already stated above, the SVD of the matrix $\overline{\mathbf{Q}} \in \mathbb{R}^{n(n+1)/2 \times N_s}$ needs be
performed only once. Its computational complexity is independent of the large dimension $N$ of the HDM and therefore is negligible
compared to the other computational costs explained next. However, the computation of all rows 
$\bar{\mathbf{h}}_i$~\eqref{eq:sol1row}, $i = 1, \dots, N$, requires the equivalent of two matrix-vector products per entry: 
its computational complexity scales as $\mathcal{O}\left(N n_{\overline{\mathbf{Q}}} \left(N_s + n(n + 1) \right) \right)$, 
where $n_{\overline{\mathbf{Q}}}$ denotes the number of singular values of $\overline{\mathbf{Q}}$.

On the other hand, the computational complexity of the SVD of the solution snapshots $\mathbf{S}$ scales as 
$\mathcal{O}\left(4 N N_s^2 + 8 N_s^3\right)$~\cite{golub1996matrix}. Noting that for large-scale HDMs
$N \gg N_s$ and $N \gg n$, assuming that $n_{\overline{\mathbf{Q}}} \sim n^2$, and assuming furthermore that enough solution snapshots are 
collected so that the least squares problems~\eqref{eq:H_opt_1row} are overdetermined -- that is, $N_s > n(n+1)/2$ --  
the complexity of the first component C1 of the total computational cost of {\bf Algorithm 1} becomes 
$\mathcal{O}\left(4 N N_s^2\right)$ and that of the second component C2 becomes $\mathcal{O}\left(2 N N_s n^2 \right)$.
Under these reasonable assumptions, computing the coefficient matrix $\overline{\mathbf{H}}$ of the quadratic term of the
quadratic approximation~\eqref{eq:quad_approx} is significantly cheaper than computing the matrix $\mathbf{V}$ needed
for both constructions of traditional and quadratic PROMs. This conclusion holds true even when the computational overhead associated with
the regularization of the least squares problems~\eqref{eq:H_opt_1row} using GCV to identify $\alpha^\star$ is accounted for,
as this computational cost is at most equal to that associated with computing the matrix $\overline{\mathbf{H}}$.

As for memory resources, the construction offline of the quadratic approximation manifold requires in principle the
storage of the matrices $\overline{\mathbf{Q}} \in \mathbb{R}^{n(n+1)/2 \times N_s}$ and 
$\mathbf{E} \in \mathbf{R}^{N\times N_s}$ -- which are needed to construct the matrix $\mathbf{H}$ -- in addition to the 
storage requirements of the traditional subspace approximation. For large-scale HDMs, the storage requirement of 
$\overline{\mathbf{Q}} \in \mathbb{R}^{n(n+1)/2 \times N_s}$ is negligible. As for the matrix $\mathbf{E}$, it does not
necessarily need to be stored in practice. Since $\mathbf{H}$ is determined row-by-row as in~\eqref{eq:sol1row}, only
$\left [\mathbf{E}\right]_{i,1}$ needs be stored for computing $\bar{\mathbf{h}}_i$, $i = 1, \dots, N$. Hence, for all 
practical purposes, only a buffer of any affordable size greater than $n(n+1)/2$ is needed for storing $\mathbf {E}$ for
the purpose of computing $\mathbf{H}$. In summary, the proposed procedure for constructing offline the quadratic approximation
manifold can be implemented in an approach that has a slighlty higher storage requirement than its counterpart for the
traditional subspace approximation.

{\it REMARK 2}. For the sake of simplicity, the notation $\mathbf{H}$ is used throughout the remainder of this paper. However,
all reported performance results and their discussion are for an implementation where $\mathbf{H}$ is represented by the $N$ 
rows $\bar{\mathbf{h}}_i$, $i = 1, \dots, N$, to avoid redundant calculations. 

\section{Impact on LSPG and ECSW}
\label{sec:IMPACT}

\subsection{Impact on the projection-based model order reduction method LSPG}
\label{sec:QUAD_LSPG}

Here, the impact of the quadratic approximation~\eqref{eq:quad_approx} on the nonlinear PMOR method LSPG outlined in the
first part of Section~\ref{sec:CONV} is highlighted. 

To this end, the quadratic approximation~\eqref{eq:quad_approx} is substituted
for the traditional affine counterpart~\eqref{eq:conventional} in the expression of the PROM~\eqref{eq:consresidual} of dimension $n$,
which leads to
\begin{equation*} 
	\mathbf{W}^{{m+1}^T} \mathbf{r}^{m+1} \left(\underbrace{\mathbf{H}\left[{\mathbf{q}^{m+1}}^{\otimes 2}\right] + \mathbf{Vq}^{m+1} + 
	\mathbf{u}_{\text{ref}}}_{{\tilde{\mathbf{u}}}^{m+1}\left(\mathbf{q}^{m+1}\right)}, t^{m+1}\right) = 0 
\end{equation*}
The first-order Taylor approximation of the above nonlinear projected residual around the $\ell$-th iterate solution $\mathbf{q}^{m+1, \ell}$ can be written as
\begin{align}
	\mathbf{W}^{{m+1, \ell+1}^T}\mathbf{r}^{m+1, \ell+1} & \left(\tilde{\mathbf{u}}^{m+1}\left(\mathbf{q}^{m+1, \ell} + \Delta \mathbf{q}^{m+1, \ell+1}\right)\right) \approx 
	\mathbf{W}^{{m+1, \ell+1}^T}\mathbf{r}^{m+1, \ell+1} \left(\tilde{\mathbf{u}}\left(\mathbf{q}^{m+1, \ell}\right)\right) \nonumber\\
	+&\,\mathbf{W}^{{m+1, \ell+1}^T}\mathbf{J}^{m+1, \ell}\left(\tilde{\mathbf{u}}\left(\mathbf{q}^{m+1, \ell}\right)\right)\pdv{\tilde{\mathbf{u}}}{\mathbf{q}}\left(\mathbf{q}^{m+1, \ell}\right)\Delta \mathbf{q}^{m+1, \ell+1}
	\label{eq:MISS0}
\end{align}
where
\begin{equation}
	\displaystyle{\frac{\partial \tilde{\mathbf{u}}}{\partial \mathbf{q}}\left(\mathbf{q}^{m+1, \ell}\right) = 
	\frac{\partial \left(\mathbf{H} \left[\mathbf{q}^{{m+1, \ell}^{\otimes 2}}\right] + \mathbf{Vq}^{m+1, \ell}\right)}{\partial \mathbf{q}}}
	\label{eq:MISS1}
\end{equation}
A nice property of a nonlinear, polynomial approximation such as that written in~\eqref{eq:polynomial} is that for any
degree $p$, it is analytically differentiable with respect to $\mathbf{q}$. In particular, for $p=2$, the derivative~\eqref{eq:MISS1} has the following analytical expression
\begin{equation}
	\displaystyle{\frac{\partial \left(\mathbf{H} \left[\mathbf{q}^{{m+1, \ell}^{\otimes 2}}\right] \right)}{\partial \mathbf{q}}}
	= \mathbf{H} \left[ \mathbf{q}^{m+1, \ell} \otimes \mathbf{I} + \mathbf{I} \otimes \mathbf{q}^{m+1, \ell}\right]
	\label{eq:MISS2}
\end{equation}
where $\mathbf{I} \in \mathbb{R}^{n \times n}$ is the identity matrix of dimension $n$. Hence, from~\eqref{eq:MISS1} and~\eqref{eq:MISS2}, it follows that
\begin{equation} 
	\displaystyle{\frac{\partial \tilde{\mathbf{u}}}{\partial \mathbf{q}}\left(\mathbf{q}^{m+1, \ell}\right) = 
	\mathbf{H} \left[ \mathbf{q}^{m+1, \ell} \otimes \mathbf{I} + \mathbf{I} \otimes \mathbf{q}^{m+1, \ell}\right] + \mathbf{V}}
	\label{eq:MISS3}
\end{equation}
and from~\eqref{eq:MISS0} and~\eqref{eq:MISS3}, it follows that the left ROB for LSPG is given in this case by
\begin{equation} 
	\mathbf{W}^{m+1, \ell+1} = \mathbf{J}^{m+1, \ell} \, \left(  \mathbf{H} \left[ \mathbf{q}^{m+1, \ell} \otimes \mathbf{I} + \mathbf{I} \otimes \mathbf{q}^{m+1, \ell}\right] + \mathbf{V} \right)
	\label{eq:WforLSPG2}
\end{equation}

The comparison of~\eqref{eq:WforLSPG2} and~\eqref{eq:WforLSPG} shows that in the case of the quadratic approximation~\eqref{eq:quad_approx}, the left ROB of LSPG contains one additional term -- namely,
$\mathbf{W}_\text{add}^{m+1, \ell+1} = \mathbf{J}^{m+1, \ell} \mathbf{H} \left[ \mathbf{q}^{m+1, \ell} \otimes \mathbf{I} + \mathbf{I} \otimes \mathbf{q}^{m+1, \ell}\right]$. Regarding this term, it is noted here
that $\left[ \mathbf{q}^{m+1, \ell} \otimes \mathbf{I} + \mathbf{I} \otimes \mathbf{q}^{m+1, \ell}\right] \in \mathbb{R}^{n(n+1)/2\times n}$ is a sparse matrix and therefore the product 
$\mathbf{H}\left[ \mathbf{q}^{m+1, \ell} \otimes \mathbf{I} + \mathbf{I} \otimes \mathbf{q}^{m+1, \ell}\right]$ should be performed using dense-sparse matrix-matrix computations. Specifically, the number of 
nonzero entries of the matrix $\left[ \mathbf{q}^{m+1, \ell} \otimes \mathbf{I} + \mathbf{I} \otimes \mathbf{q}^{m+1, \ell}\right]$ grows as $n^2$ and the evaluation at each Gauss-Newton iteration of the product of 
$\mathbf{H}$ and this sparse matrix requires $\mathcal{O}\left(N n^2\right)$ operations (in the absence of hyperreduction and when redundant computations are avoided). 

In summary, in the case of the quadratic PROM, the construction of $\mathbf{W}^{m+1, \ell+1}$ requires $\mathcal{O}(2N^3n)$ operations whereas
in the case of the traditional PROM, it requires $\mathcal{O}(N^3n)$ operations. Hyperreduction however eliminates the dependence on $N$ from both aforementioned computational complexities
which become quite reasonable.

From a storage perspective, it follows from~\eqref{eq:MISS1} and~\eqref{eq:MISS2} that the construction of the left ROB of LSPG~\eqref{eq:WforLSPG2} associated with the quadratic PROM requires the 
additional storage of the matrix $\mathbf{H}_{\mathbf{q}} =  \mathbf{H} \left[ \mathbf{q}^{m+1, \ell} \otimes \mathbf{I} + \mathbf{I} \otimes \mathbf{q}^{m+1, \ell}\right] \in \mathbb{R}^{N \times n}$.

\subsection{Impact on the hyperreduction method ECSW}
\label{sec:ECSW_QUAD_IMPACT}

With regard to the impact of the quadratic approximation~\eqref{eq:quad_approx} on the hyperreduction method ECSW outlined in the second part of Section~\ref{sec:CONV}, two comments are noteworthy: one 
pertaining to the construction of the non negative least squares (NNLS) problem~\eqref{eq:nnls_ecsw} and its early termination criterion~\eqref{eq:nnls_ecsw_early_termination}; and one pertaining to the 
size of the resulting reduced mesh.

As highlighted in~\eqref{eq:training} and the first bullet below that equation, it is important to perform the ECSW training
for the PROM predictions and not those of its underlying HDM. In the case of the traditional affine approximation, this
is achieved by projecting the solution snapshots collected in the matrix $\mathbf S$ on the right ROB $\mathbf{V}$; and
constructing the matrix $\mathbf{C}$ and vector $\mathbf{d}$ defining the least squares problem~\eqref{eq:nnls_ecsw} and its
early termination criterion~\eqref{eq:nnls_ecsw_early_termination} using the projected snapshots. In the case of the
proposed quadratic approximation~\eqref{eq:quad_approx}, the identification of the vector of generalized coordinates
$\mathbf{q}_i$ associated with a solution snapshot $\mathbf{u}_i$ can no longer be performed via projection onto a ROB.
Instead, it requires the solution of a nonlinear problem of the form
\begin{equation*} 
	\boldsymbol{\delta}_i(\mathbf{q}_i) = \mathbf{H}\left[\mathbf{q}_i^{\otimes 2} \right] + \mathbf{Vq}_i + \mathbf{u}_{\text{ref}} - \mathbf{u}_i = \mathbf{0}
\end{equation*}
using, for example, yet another Gauss-Newton procedure that can be summarized as follows
\begin{align*}
	\mathbf{q}_i^{0} & = \mathbf{V}^T \mathbf{u}_i \\ 
	\mathbf{q}_i^{\ell+1} & = \mathbf{q}_i^{\ell} - \left(\pdv{\boldsymbol{\delta}_i}{\mathbf{q}}\left(\mathbf{q}_i^{\ell}\right)\right)^+
	\boldsymbol{\delta}_i\left(\mathbf{q}_i^{\ell}\right)
\end{align*}
where the superscript $\ell$ designates here too the $\ell$-th iteration and the superscript $+$ designates the Moore-Penrose inverse. 
Hence, for the same number $N_h$ of training solution snapshots, the computational cost associated with the setup of ECSW 
in the case of the proposed quadratic approximation manifold is higher than in the case of the traditional affine approximation.
However, as explained below, this potential cost increase is amply justified and rewarded.

As justified in Section~\ref{sec:HEURISTIC}, given a nonlinear, convection-dominated problem, a quadratic PROM can
be expected to deliver a similar level of accuracy as a traditional PROM using however a smaller (and potentially
much smaller) dimension $n$. Considering this and the fact that the number of cubature points required for approximating a 
$d$-dimensional integral function with $p$ cubature points along each dimension grows as $p^d$, it follows that
one can reasonably expect ECSW to deliver in the context of the proposed quadratic approximation manifold a smaller to
much smaller reduced mesh than in the context of the traditional subspace approximation. Hence, the hyperreduction
using ECSW of a quadratic PROM can be expected to deliver an even better computational efficiency than that of a
traditional PROM. The numerical results reported in Section~\ref{sec:APP} for the prediction of the highly nonlinear, 
convection-dominated, turbulent flow around and in the wake of the Ahmed body support this conclusion.

\section{Application}
\label{sec:APP}

The proposed quadratic approximation manifold was implemented in the compressible flow solver AERO-F, which is equipped with low-Mach preconditioning~\cite{turkel1993review}, and added to its
nonlinear model reduction capabilities. In order to showcase the performance of this approximation manifold, it is applied here to the acceleration of the simulation of the Ahmed body turbulent wake 
flow~\cite{ahmed1984some}, which is a popular CFD benchmark problem in the automotive industry. The Ahmed body geometry can be described as an extruded rectangle, where the front 
(facing the free-stream velocity) is rounded to promote a separation-free flow, the middle is a rectangular parallelepiped to promote a uniform flow, and the rear is downward slanted at 
varying angles to promote the generation of the wake flow of interest. Motivating factors for this geometry were the understanding of the dependence of the wake and 
experimentally measured drag on the slant angle, for the purpose of design optimization for fuel economy; and the generation of validation data for computational models~\cite{ahmed1984some}. 
Even in the absence of a parametric setting -- except for time -- this problem is suitable for assessing the performance of the quadratic approximation manifold proposed in this paper, because the flow 
is modeled here using the detached-eddy simulation (DES) approach\footnote{DES is a modification of a RANS model in which the model switches to a subgrid scale formulation in regions of the flow that 
are fine enough for large eddy simulation (LES) computations}~\cite{strelets2001detached}. The unsteadiness of the wake flow provides in this case a sufficiently rich solution 
for observing the Kolmogorov barrier without resorting to further parameterization. Indeed, as already stated in Section~\ref{sec:KB}, it was reported in~\cite{grimberg2021mesh} that for this unsteady 
viscous flow problem, a global ROB of dimension roughly equal to 600 is needed to construct a PROM or HPROM that is reasonably accurate with respect to standard aerodynamic QoIs such as the
time-history of the drag coefficient~\cite{grimberg2021mesh}.

Troughout this section, all PROMs are constructed using LSPG and hyperreduced using ECSW. A traditional LSPG PROM -- that is, an LSPG PROM constructed using the traditional subspace 
approximation~\eqref{eq:conventional} -- is simply referred to as a PROM -- and an HPROM after hyperreduction; and any counterpart quadratic PROM constructed using LSPG and the proposed quadratic 
approximation~\eqref{eq:quad_approx} is referred to as a QPROM -- and HQPROM after hyperreduction.

All computations reported in this section are performed using AERO-F; and double-precision arithmetic on a Linux cluster where each node is configured with two Intel Xeon Gold 5118 processors clocked
at 2.3 GHz and a total of 24 cores, and 192 GB of memory. Specifically, any HDM, PROM, QROM, HPROM, or HQPROM is constructed using a third-order spatial approximation of the convective fluxes;
a second-order spatial approximation of the viscous fluxes; and a second-order time-discretization. Each unsteady simulation is initialized using the solution of its quasi-steady-state counterpart.

For this application, the accuracy of each constructed HPROM or HQPROM is assessed by focusing on a given QoI and measuring the relative error of the prediction delivered by the HPROM or HQPROM
with respect to that obtained using the underlying HDM, as follows
\begin{equation}
	\mathbb{RE}_{\text{QoI}} = \frac{\sqrt{\sum\limits_{t \in \mathcal{T}}\left(\widetilde{QoI}(t) - QoI(t)\right)^2}}{\sqrt{\sum\limits_{t \in \mathcal{T}} QoI(t)^2}}
	\label{eq:relative_error}
\end{equation}
where $\widetilde{QoI}(t)$ designates the QoI based on the reconstructed solution delivered by the HPROM or HQPROM, $QoI(t)$ denotes its counterpart value based on the underlying HDM, 
and \break $\mathcal{T} = \left\{ t \in \{0, \Delta s, 2 \Delta s, \dots\} : t \leq 2\times 10^{-1}~\hbox{s} \right\}$ denotes the set of time-stamps associated with the collected, HDM-based,
solution snapshots.

\subsection{High-dimensional CFD model}
\label{sec:HDM}

The slant angle is fixed at 20$^\circ$, the free-stream velocity is set to $v_\infty = 60$ m/s, and the free-stream angle of attack is set to $0^{\circ}$. 
In this case, the Reynolds number based on the body length as the length of reference is $Re = 4.29 \times 10^6$. The symmetry of the body with respect to the plane $y = 0$ m is accounted for: 
therefore, the computational fluid domain is defined as a rectangular parallelepiped aligned on one end with the plane of symmetry. It is discretized by a body-fitted unstructured mesh with 17,017,090 
tetrahedra and 2,890,434 vertices with $d_e = 6$ unknowns per vertex, which leads to a CFD HDM of dimension $N = 17,342,604$. Air is modeled as a perfect gas, adiabatic boundary conditions are applied 
to the wall boundary of the computational fluid domain, and the DES model is equipped with Reichardt's law of the wall. The simulation time-interval is set to $[0, 2\times 10^{-1}]$ s and 
time-discretization is performed using the second-order three-point backwards difference scheme with the fixed dimensional time-step $\Delta t = 8 \times 10^{-5}$ s.

Using 10 nodes of the aforementioned Linux cluster and thus a total number of 240 cores, the HDM-based simulation requires 15.1 h wall clock time, which corresponds to $3.624 \times 10^3$ h CPU time. 
It predicts a time-averaged drag coefficient $\overline{c}_D = 0.262$, where averaging is performed over the time-subinterval $[5 \times 10^{-2}, 2 \times 10^{-1}]$ s and $t = 5 \times 10^{-2}$ s
corresponds to the time-instance after which all transient effects of the simulation have vanished. For reference, the experimental value of the time-averaged drag coefficient reported 
in~\cite{ahmed1984some} is $\overline{c}_D^{\ \text{exp}} = 0.255$. Therefore for this QoI, the HDM is predictive with a relative error of 2.75\%.

\subsection{Nonlinear LSPG reduced-order models}

For the purpose of PMOR, $N_s = 1,251$ solution snapshots are collected during the HDM-based simulation -- specifically, at every other time-step of the simulation. Using these snapshots,
a PROM and a QPROM are constructed as described in Section~\ref{sec:CONV} and Section~\ref{sec:QM}, respectively. 

In the case of a PROM, the dimension $n$ of the ROB is determined using the singular value energy threshold $\varepsilon_{\mathbf{S}} = 10^{-4}$ (see~\eqref{eq:singular_value_energy_criteria}) -- or 
equivalently, $1 - \varepsilon_{\mathbf{S}} = 99.99\%$ -- which results in the relatively large value of $n = 627$. 

In the case of an HQPROM, $n$ is determined using the heuristic approach presented in Section~\ref{sec:HEURISTIC}. In this case, item I3 in that section leads to
$n_{\text{qua}}^\prime = \left(\sqrt{9 + 8\times 627} - 3\right)/2 \approx 34$. 
Next, setting in item I4 $\zeta = 0.15$ to account for vulnerability to overregularization leads to $n_{\text{qua}} = (1 + 0.15)n_{\text{qua}}^\prime \approx 39$. Then, applying~\eqref{eq:CONS} 
to satisfy the constraints recalled in item I1 gives $n = \min \left(39, \left(\sqrt{1 +8 \times 1,251} - 1\right)/2 \right) = \min \left(39, 44\right) = 39$. Applying GCV with $\omega = 0.1$ yields 
$\alpha^\star / \sigma_{\overline{\mathbf{Q}}, 1} = 2.8 \times 10^{-5}$, where $\alpha^\star$ is reported as a multiplier of the highest singular value of $\overline{\mathbf{Q}}$ (see~\eqref{eq:sol1row})
to highlight the strength of the regularization.

\subsection{Nonlinear LSPG hyperreduced reduced-order models and accuracy results}
\label{sec:acc}

Next, ECSW is configured with the parallel variant of the NNLS algorithm of Lawson and Hanson~\cite{lawson1995solving} described in~\cite{chapman2017accelerated} 
to solve the convex optimization problem~\eqref{eq:nnls_ecsw} equipped with the early termination criterion~\eqref{eq:nnls_ecsw_early_termination} and $\tau = 10^{-2}$. Its application to the CFD
mesh described in Section~\ref{sec:HDM} produces for each ROB underlying the construction of the PROM or QPROM the reduced mesh needed for building the HPROM or HQPROM, respectively.
The statistics of both reduced meshes computed using all 240 cores are summarized in Table~\ref{tab:ecsw} below. In this table, the reader can observe that using the same value of the training tolerance 
$\tau$, ECSW produces an order of magnitude smaller reduced mesh when training is performed in the context of a QPROM, than when it is performed in that of a PROM. This result is consistent with the 
expectation set in Section~\ref{sec:ECSW_QUAD_IMPACT}, based on the fact that the number of cubature points required for approximating a $d$-dimensional integral function with $p$ cubature points along 
each dimension grows as $p^d$ (here, $n$ plays the role of $d$). Note also that ECSW constructs the reduced mesh for the HQPROM in 1.84 mn -- that is, 259 times faster than for the HPROM, underscoring a 
first computational advantage of the quadratic approximation manifold proposed in this paper.

\begin{table}[ht]
\centering
\begin{tabular}{c|c|c|c|c|c}
Computational model & $\alpha^{\star} / \sigma_{\overline{\mathbf{Q}}, 1}$ & $n$ & $n_e$ & $n_e / N_e$ (\%) & NNLS solve time (h) \\ \hline
PROM $\to$ HPROM & -- & 627 & 7,389 & 0.26 & 7.94\\ \hline
QPROM $\to$ HQPROM & $2.8 \times 10^{-5}$ & 39 & 544 & 0.019 & 0.0306
\end{tabular}
\caption{Application of ECSW configured with the parallel NNLS algorithm and $\tau = 10^{-2}$ to the reduction on 240 cores of the CFD mesh containing $N_e = 2,890,434$ vertices: case of a traditional 
	PROM; and case of a quadratic PROM.}
\label{tab:ecsw}
\end{table}

Then, the flow simulation is repeated twice on 8 cores of a single node of the same Linux cluster: once using the HPROM; and once using the HQPROM.

\begin{figure}[ht]
\centering
\begin{subfigure}[b]{\textwidth}
\centering
\includegraphics[trim=0 400 170 400,width=\textwidth]{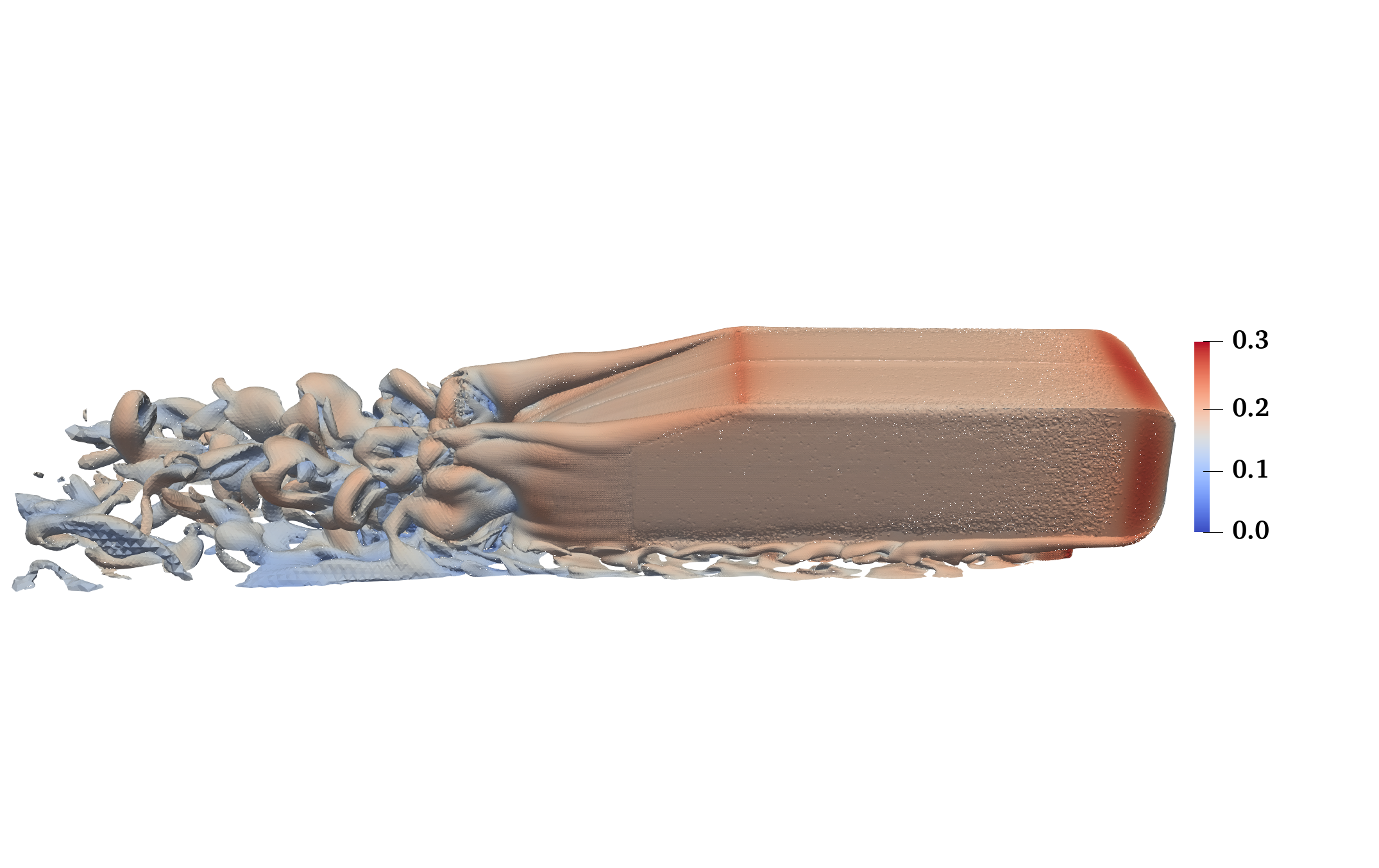}
\caption{}
\end{subfigure}
\begin{subfigure}[b]{\textwidth}
\centering
\includegraphics[trim=0 400 170 400,width=\textwidth]{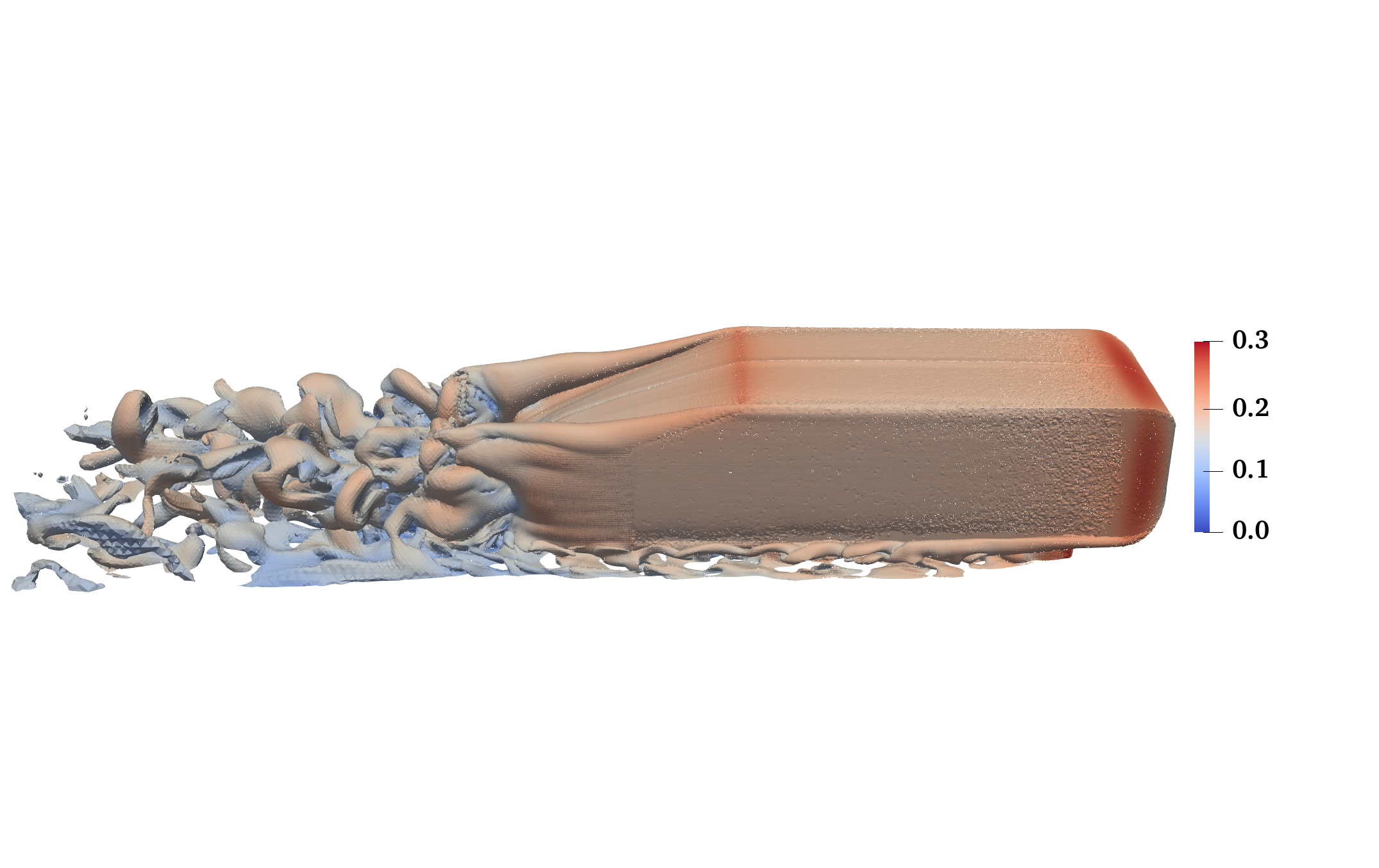}
\caption{}
\end{subfigure}

\begin{subfigure}[b]{\textwidth}
\centering
\includegraphics[trim=0 400 170 400,width=\textwidth]{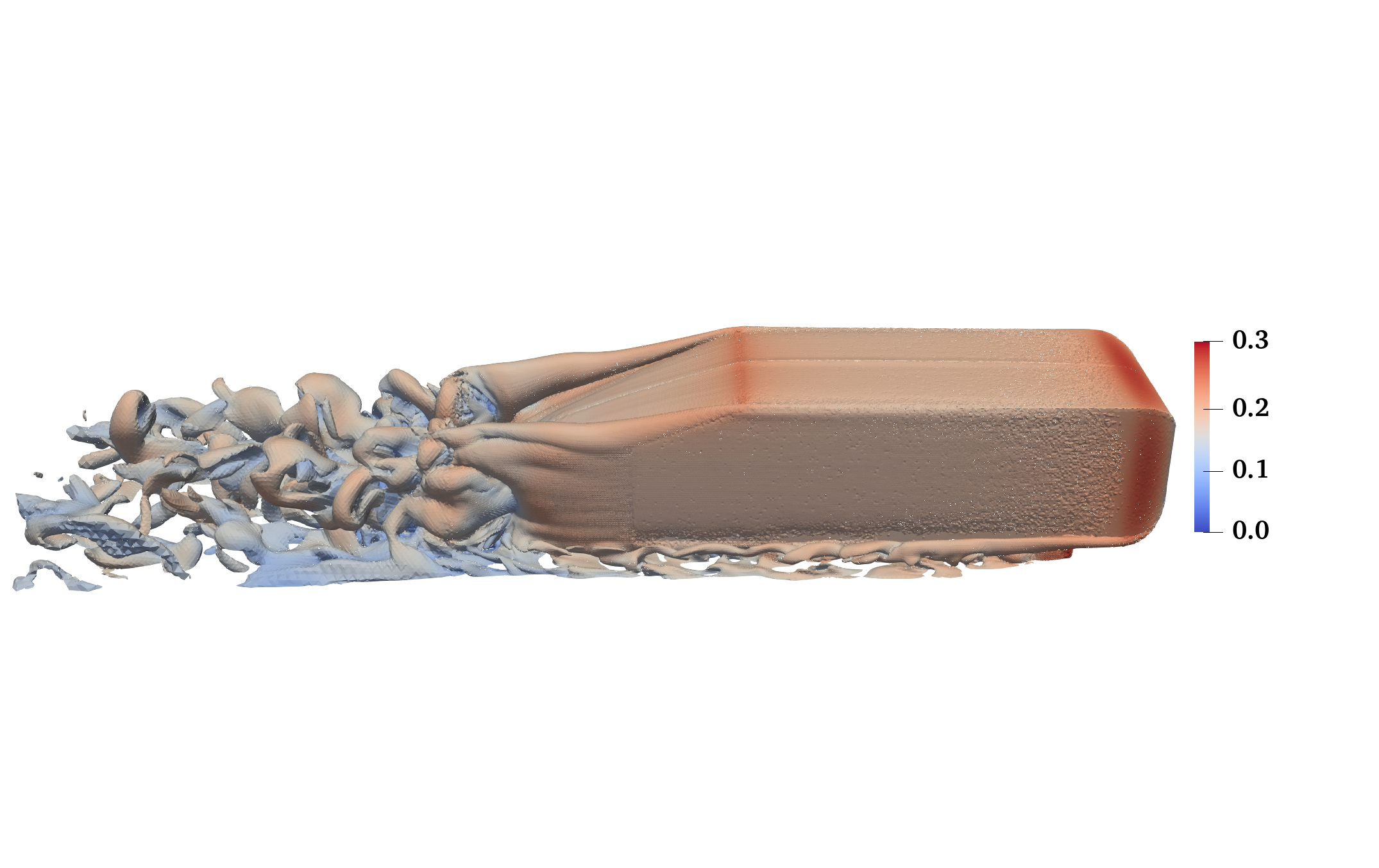}
\caption{}
\end{subfigure}
\caption{Ahmed body turbulent wake flow problem -- Iso-vorticity contours colored by the local Mach number and computed at $t = 2\times 10^{-1}$ s by postprocessing the solutions
	obtained using: (a) the HDM; (b) the HPROM ($n = 627$); and (c) the HQPROM ($n = 39$).}

\label{fig:flow_viz}
\end{figure}

Figure~\ref{fig:flow_viz} visualizes the iso-vorticity contours of the flow computed at $t = 2 \times 10^{-1}$ s (the end of the simulation time-interval) by all three considered computational models.
It shows that the three computed fields are largely indistinguishable. 

\clearpage

\clearpage

\begin{figure}[ht]
\centering
\begin{subfigure}[b]{0.75\textwidth}
\centering
\includegraphics[width=\textwidth]{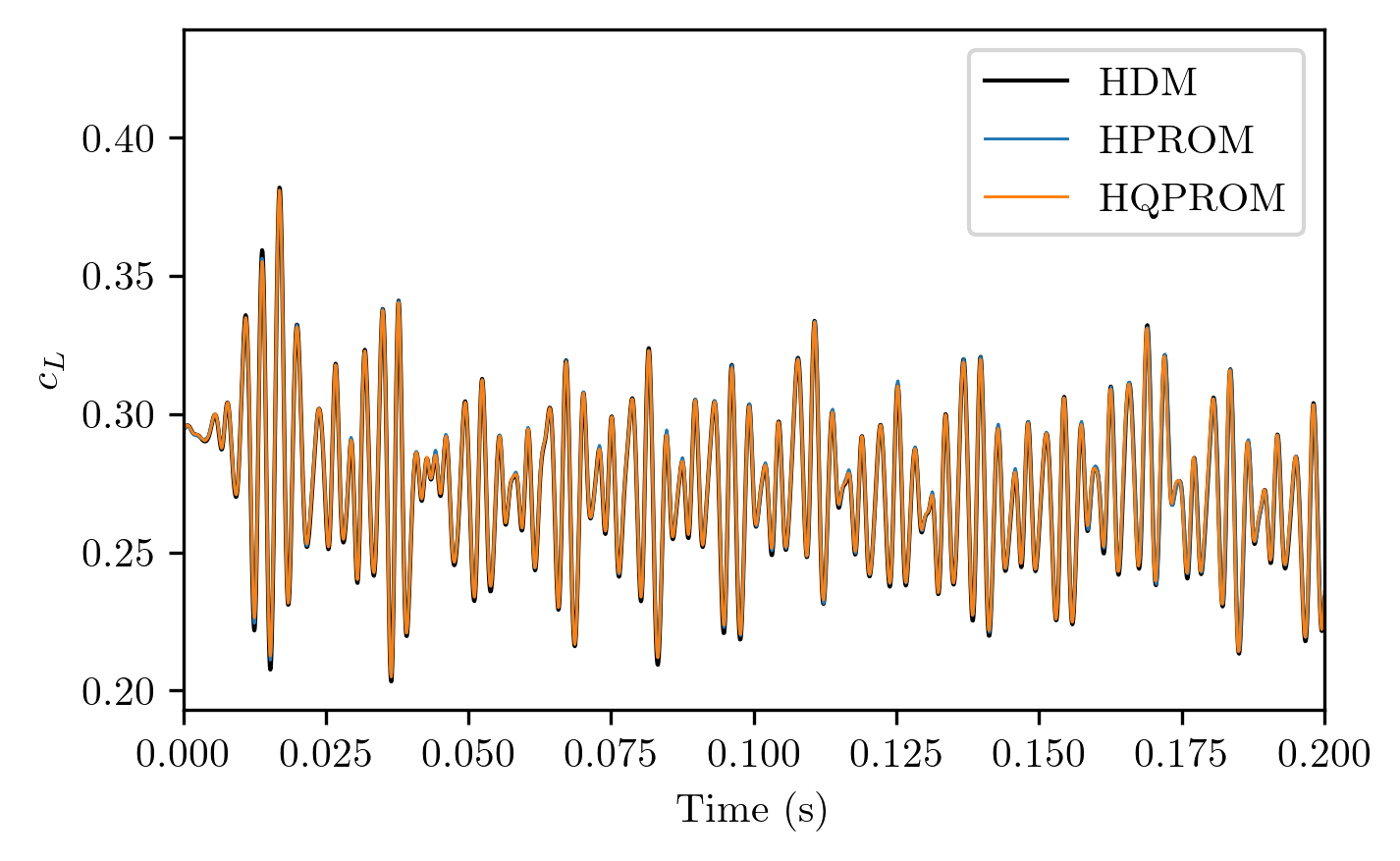}
\caption{}
\end{subfigure}
\begin{subfigure}[b]{0.75\textwidth}
\centering
\includegraphics[width=\textwidth]{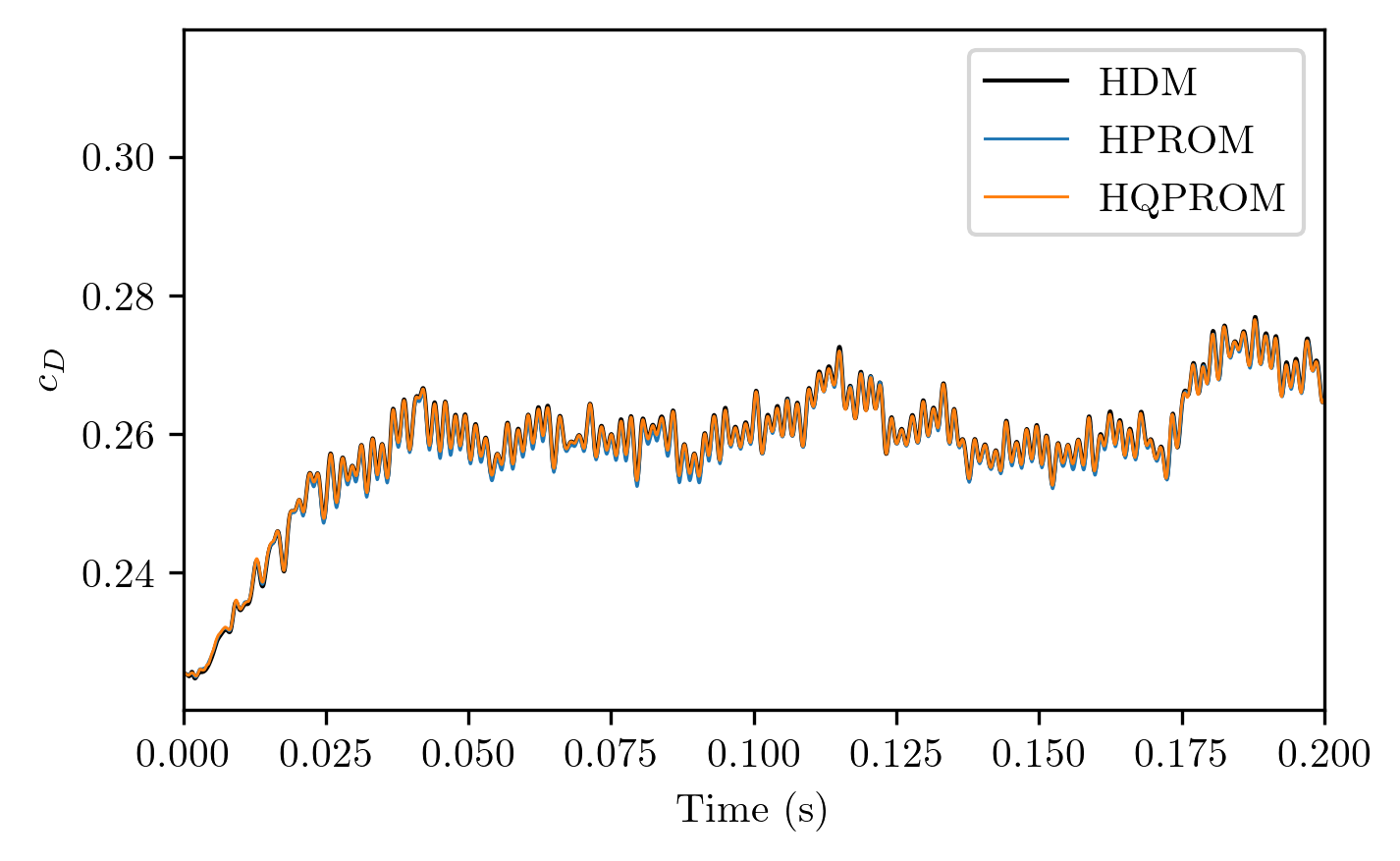}
\caption{}
\end{subfigure}
	\caption{Ahmed body turbulent wake flow problem -- Time-histories of the lift (a) and drag (b) coefficients predicted using: the HDM; the HPROM; and the HQPROM.}
\label{fig:QoI-cd}
\end{figure}

\begin{figure}[ht]
\centering
\begin{subfigure}[b]{0.75\textwidth}
\centering
\includegraphics[width=\textwidth]{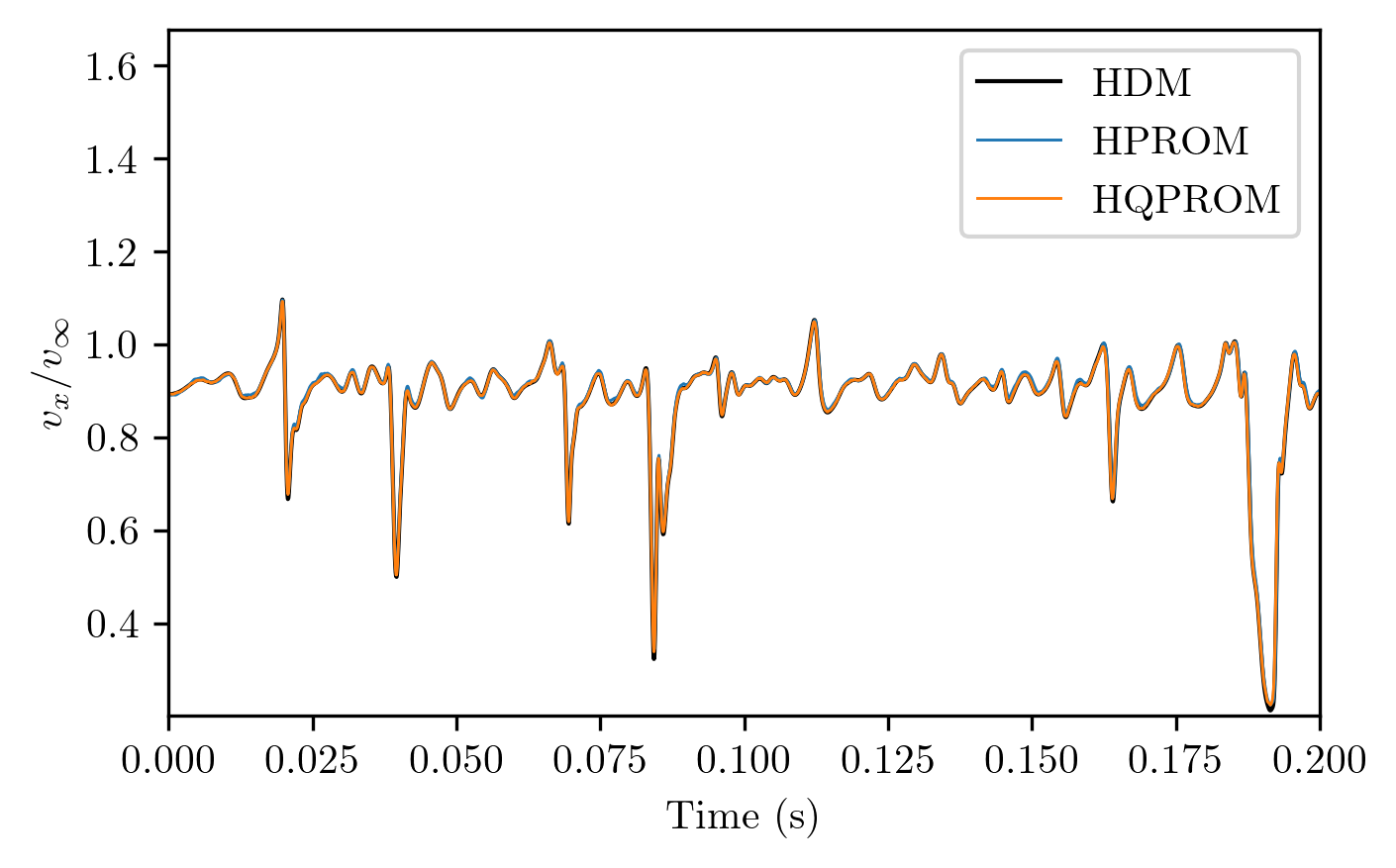}
\caption{}
\end{subfigure}
\begin{subfigure}[b]{0.75\textwidth}
\centering
\includegraphics[width=\textwidth]{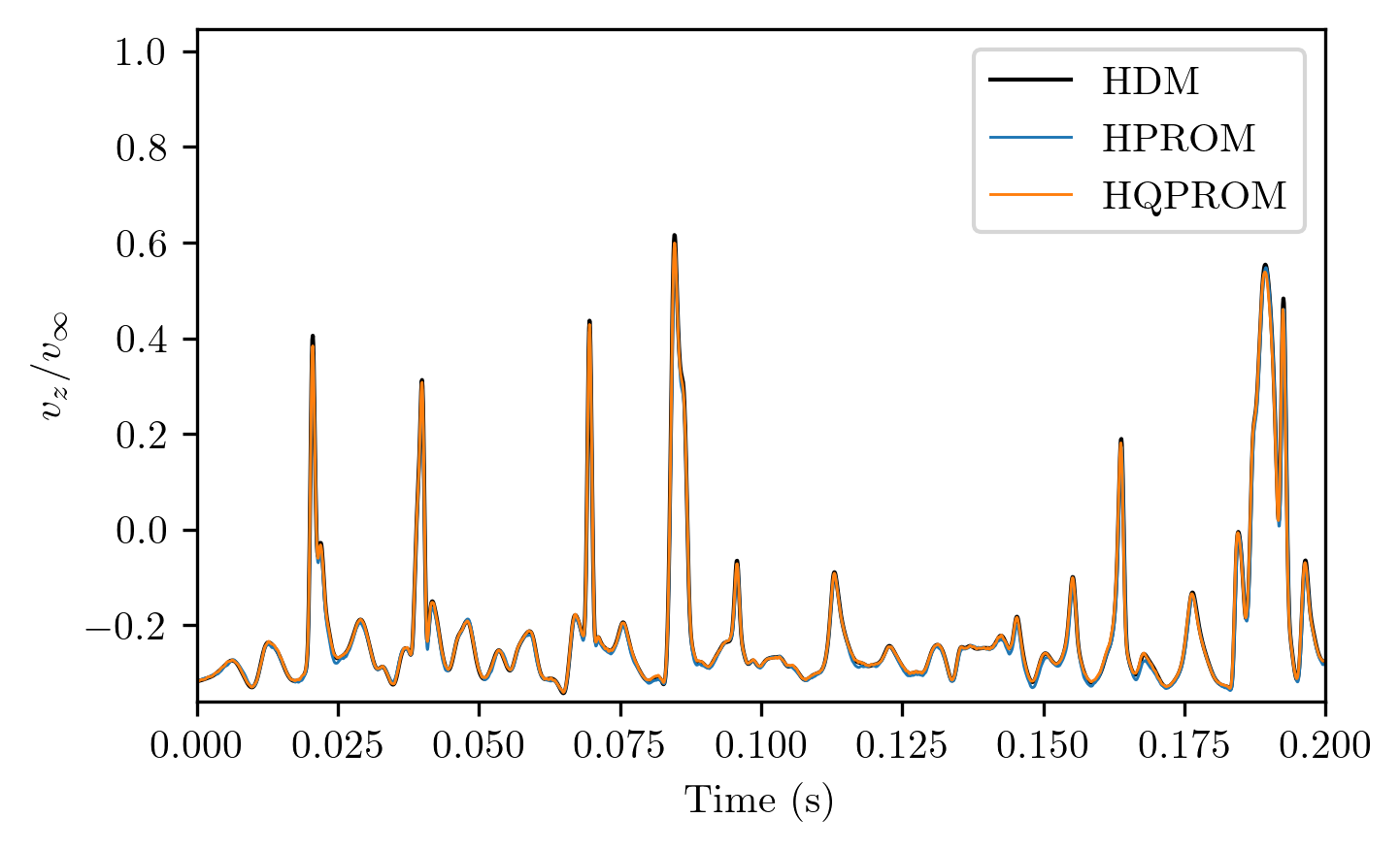}
\caption{}
\end{subfigure}
	\caption{Ahmed body turbulent wake flow problem -- Time-histories of $v_x/v_{\infty}$ (a) and $v_z/v_{\infty}$ (b) predicted using: the HDM; the HPROM; and the HQPROM.}
\label{fig:QoI-vel}
\end{figure}

Whereas Figure~\ref{fig:flow_viz} focuses on a spatio-temporal QoI, each of Figure~\ref{fig:QoI-cd} and Figure~\ref{fig:QoI-vel} focuses on two scalar but time-dependent QoIs: the time-histories of the 
lift and drag coefficients (Figure~\ref{fig:QoI-cd}), which are integral quantities; and those of the $x$- and $z$-components of the flow velocity at a probe located along the wake of the flow, 
normalized by the magnitude of the free-stream velocity $v_{\infty}$ (Figure~\ref{fig:QoI-vel}), which are local QoIs (note that the $x$ direction is in this case the main direction of the flow). In both 
figures, the reader can observe that each of the HPROM and HQPROM delivers an excellent accuracy. This conclusion is supported by the computed relative errors reported in 
Table~\ref{table:relative-errors}, which also reveal that strictly speaking, the HQPROM delivers for the considered QoIs slightly more accurate results than the HPROM.

\clearpage

\clearpage

\begin{table}[h!]
\centering
\begin{tabular}{c|c|c|c|c|c|c}
Computational model & $\alpha^{\star} / \sigma_{\overline{\mathbf{Q}}, 1}$ & $n$ & $\mathbb{RE}_{c_D}$ (\%) & $\mathbb{RE}_{c_L}$ (\%) & $\mathbb{RE}_{v_x}$ (\%) & $\mathbb{RE}_{v_z}$ (\%) \\ \hline
HPROM & -- & 627 & 0.24 & 0.77 & 0.83 & 3.95 \\ \hline
HQPROM & $2.8 \times 10^{-5}$ & 39 & 0.10 & 0.71 & 0.54 & 2.66
\end{tabular}
\caption{Ahmed body turbulent wake flow problem -- Relative errors associated with the predictions of the considered QoIs delivered by the HPROM and HQPROM.}
\label{table:relative-errors}
\end{table}

\subsection{Performance results}
\label{sec:AHMED_RESULTS}

Table~\ref{table:offline-compute-time} reports for each of the constructed HPROM and HQPROM the wall clock timings on 240 cores of the Linux cluster associated with the computation of
the ROB $\mathbf{V}$, that of the matrix $\mathbf{H}$ (using {\bf Algorithm 1} with $\omega = 0.1$ and $\zeta = 0.15$) in the case of the HQPROM, and the entire offline computations. The results reported in this table
show that:
\begin{itemize}
	\item The cost associated with the computation of the ROB $\mathbf{V}$ is almost identical in both cases: this is expected since the two ROBs share the same snapshot matrix and its SVD, and 
		differ only in the truncation of the singular values and therefore in size.
	\item In the case of the HQPROM, the cost associated with the computation of the matrix $\mathbf{H}$ is smaller than that associated with the computation of the ROB $\mathbf{V}$. As anticipated
		in Section~\ref{sec:QM}, this is largely due to the embarrassingly parallel method for computing $\mathbf{H}$ which, for this representative application, scales better on 240 cores 
		than any parallel SVD applied to the computation of $\mathbf{V}$. It is noted that strictly speaking, the performance of {\bf Algorithm 1} depends on the inputted value of $\omega$
		(see {\bf Algorithm}~\ref{alg:calc_H}). Everything else being equal, the upper bound of the computational cost of this algorithm is reached for $\omega = 1$ (as in this case the largest 
		number $n_{\text{smp}}$ of trial values of $\alpha$ is explored by GCV), which for this application leads to a similar value of $\alpha^{\star}$ and to the wall clock 
		timing of $7.03 \times 10^2$ s. Hence, even in this worst case scenario, computing $\mathbf{H}$ for this application is faster than computing the ROB $\mathbf{V}$.
	\item The total offline cost, which, in addition to the costs mentioned above -- as applicable -- includes the cost of constructing a reduced mesh using ECSW and I/O as well as 
		interprocessor communication as needed, is five times smaller in the case of the construction of the HQPROM. This is largely due to the much faster computation by ECSW of a reduced mesh
		in the case of an HQPROM, as explained in Section~\ref{sec:ECSW_QUAD_IMPACT} and shown in Table~\ref{tab:ecsw}.
\end{itemize}

\begin{table}[h!]
\centering
\begin{tabular}{c|c|c|c|c|c}
Computational & $\alpha^{\star} / \sigma_{\overline{\mathbf{Q}}, 1}$ & $n$ & Wall clock time (s) & Wall clock time (s) & Wall clock time (s)\\
model         &                                                      &     & ($\mathbf{V}$)      & ($\mathbf{H}$)      & (total offline) \\ \hline
HPROM & -- & 627 & $1.01\times 10^3$ & -- & $3.25 \times 10^4$ \\ \hline
HQPROM & $2.8 \times 10^{-5}$ & 39 & $1.02\times 10^3$ & $5.75 \times 10^2$ & $6.23 \times 10^3$ \\
\end{tabular}
\caption{Ahmed body turbulent wake flow problem -- offline wall clock timings on 240 cores of a Linux cluster.}
\label{table:offline-compute-time}
\end{table}

Table~\ref{table:speedup-factors} reports for each constructed reduced-order model its online execution time on 8 cores of the same Linux cluster and its speed-up factors with respect to the execution 
time of the HDM given in Section~\ref{sec:HDM}. Because the HDM-based simulation is performed on 240 cores, two speed-up factors are reported for each constructed reduced-order model: one measured using 
wall clock timings; and another one measured using CPU timings. The first speed-up factor is most important for time-critical applications. The second one is relevant for resource- as well as 
time-critical applications. In this table, the reader can observe that the HQPROM is more than 32 times faster than the HPROM (while, as shown in Section~\ref{sec:acc}, delivering similar if not
better accuracy). Consequently, while the HPROM delivers a speed-up factor of 4 only for wall clock time, the HQPROM delivers a speed-up factor of 131 by this measure. Using the CPU time measure however,
the HPROM and HQPROM deliver speed-up factors of two orders and three orders of magnitude, respectively.

\begin{table}[ht]
\centering
\begin{tabular}{c|c|c|c|c|c}
Computational model & $\alpha^{\star} / \sigma_{\overline{\mathbf{Q}}, 1}$ & $n$ & Wall clock time (s) & Speed-up factor                & Speed-up factor         \\ 
(number of cores)   &                                              &     &                     & (wall clock time)              & (CPU time)              \\ \hline
HDM (240)   & -- & --  & $5.45 \times 10^4$ & -- & -- \\ \hline
HPROM (8) & -- & 627 & $1.35\times 10^4$ & $ 4.04 \times 10^0$ & $1.21\times 10^2$ \\ \hline
HQPROM (8) & $2.8 \times 10^{-5}$ & 39 & $4.15 \times 10^{2}$ &$1.31 \times 10^2$ & $3.94\times 10^3$
\end{tabular}
\caption{Ahmed body turbulent wake flow problem -- online timings and speed-up factors.}
\label{table:speedup-factors}
\end{table}

\section{Conclusions}
\label{sec:CONC}

Based on the results of its application to the solution of the Ahmed body turbulent wake flow problem using a detached-eddy simulation model, whose solution is sufficiently rich for observing the 
Kolmogorov barrier without resorting to parameterization beyond time, the quadratic approximation manifold proposed in this paper for constructing a quadratic projection-based reduced-order model
(QPROM) can be characterized as follows. It is a departure from the traditional approach for constructing a PROM based on the affine subspace approximation. It delivers a QPROM that, after
hyperreduction, achieves the same if not better solution accuracy than a traditional hyperreduced PROM (HPROM), using however: an order of magnitude smaller dimension; and an order of magnitude smaller 
reduced mesh for hyperreduction approximations. For the same training tolerance, the energy-conserving sampling and weighting (ECSW) hyperreduction method computes a reduced mesh for a QPROM two 
orders of magnitude faster than for a traditional PROM. For this reason, for this application and a fixed level of accuracy, the total offline computational cost associated with the construction of a hyperreduced QPROM (HQPROM) 
is five times smaller than that associated with the construction of a traditional HPROM, which greatly enhances the practicality of the proposed quadratic approximation approach. Most importantly, for the same level of accuracy,
the HQPROM performs all online computations more than an order of magnitude faster than the HPROM. All these results demonstrate the potential of HQPROMs for mitigating the Kolmogorov barrier to model 
reduction for realistic, nonlinear, convection-dominated transport problems.

In principle, the quadratic manifold approximation proposed in this paper is extendible to a piece-wise quadratic formulation using the method of most-appropriate local right ROBs described 
in~\cite{amsallem2012nonlinear} -- which by itself, is another approach for mitigating the Kolmogorov barrier to model reduction that was shown in~\cite{washabaugh2012nonlinear, grimberg2021mesh}
to be successful in three dimensions for realistic, three-dimensional, nonlinear, convection-dominated turbulent flow problems. Hence, one may reasonably expect a piece-wise quadratic manifold 
approximation to be an even more effective approach for addressing the Kolmogorov barrier for nonlinear model reduction. The development of such an approximation is an on-going work that the authors 
hope to report on soon, separately.

\section{Acknowledgments}

The authors acknowledge the support by the Air Force Office of Scientific Research under Grant FA9550-20-1-0358 and Grant FA9550-22-1-0004.

\section{References}

\bibliographystyle{unsrt}
\bibliography{refs}
\end{document}